\documentclass[pra,
a4paper,
showpacs,
twocolumn,
superscriptaddress]{revtex4-1}
\usepackage{amssymb}
\usepackage{amsmath}
\usepackage{amsfonts}
\usepackage{braket}
\usepackage{graphicx}
\usepackage{bm}
\usepackage{color}
\usepackage{multirow}
\usepackage{natbib}
\usepackage[normalem]{ulem}
\usepackage{verbatim}
\usepackage{dcolumn}
\usepackage{ulem}
\usepackage{float}

\def \be {\begin{equation}}
\def \ee {\end{equation}}

\newcommand{\aleq}[1]{
\begin{equation}
    \begin{aligned}
    #1
    \end{aligned}
\end{equation}
}

\begin{document}
\title
{ 
Tunable helical crystals 
}

\author{R.\,A.~Niyazov}
\address{NRC ``Kurchatov Institute'', Petersburg Nuclear Physics Institute, Gatchina 188300, Russia}
\address{Ioffe Institute,
194021 St.~Petersburg, Russia}
\author{D.\,N.~Aristov}
\address{NRC ``Kurchatov Institute'', Petersburg Nuclear Physics Institute, Gatchina 188300, Russia}
\address{Ioffe Institute,
194021 St.~Petersburg, Russia}
\address{Department of Physics, St. Petersburg State University, St. Petersburg 199034, Russia}
\author{V.\,Yu.~Kachorovskii }
\address{Ioffe Institute,
194021 St.~Petersburg, Russia}

\begin{abstract}
We consider a superlattice  formed by  tunnel-connected  identical holes,  periodically placed in a  two-dimensional topological insulator.   
We study tunneling transport through helical edges of these holes and demonstrate that the band structure  of such  helical crystal can be  controlled by both gate electrodes and external  magnetic filed. For   integer and half-integer values of   dimensionless magnetic flux  through the holes, 
the spectrum possesses  {Dirac} points  whose positions and velocities  can be tuned by gates.    
The deviation of  magnetic  flux from these special values by $\delta \phi$ makes the Dirac cones massive, with the gap value $\Delta \propto |\delta \phi|$. 
At certain gate-dependent  values of  $\delta \phi$   
 different Dirac points  converge to a double Dirac point and then disappear with further increase of $\delta \phi.$    
Interaction between carriers may lead to strong renormalization of  parameters $\alpha$ and $\beta$ 
controlling  total  tunnel coupling  between holes  
and   spin flip tunneling processes, respectively.     
We plot the renormalization flow  in the plane $(\alpha,\beta)$  and   demonstrate   
{\it multicritical} behavior  of the crystal---there is a multicritical 
fully unstable  fixed point   separating  three different phases: independent rings, independent shoulders,  and perfect spin-flip channels.   
We also find that defects in the crystal may lead to a formation of topologically protected  qubits which are not destroyed by temperature  
and can be also manipulated  both by gates and by magnetic field.  
The possibility of purely  electrical high-temperature control of the qubits opens a wide avenue for applications in the area of quantum computing.             
\end{abstract}

\maketitle

\section{Introduction}

One of the hot topics  actively discussed in the last decade is the electrical and optical properties of topological insulators, i.e. materials that are insulating  in the bulk and have conducting states at the boundary~\cite{Bernevig2013,Hasan2010,Qi2011}. In particular, in two-dimensional (2D)  topological insulators, nontrivial bulk band topology results in appearance of helical one-dimensional (1D) states that carry a current along the sample edges without dissipation.  The electron propagation in such  1D channels is characterized by  a certain  helicity, i.e. electrons with opposite spin propagate in opposite directions. As a remarkable consequence, scattering by non-magnetic  impurities is forbidden, and precisely because of this property, there is no dissipation in such channels.

The most well-known implementation of 2D  topological insulators is quantum wells in HgTe-based compounds,  where topological properties were predicted theoretically  \cite{Kane2005,predicted} and  confirmed  by a series of experiments including measurements of conductance of the  edge states \cite{confirmed} and  experimental evidence  of the non-local transport   \cite{Roth2009,Gusev2011,Brune2012,Kononov2015}.

 One of the new and promising areas of research   is the interferometry based on helical 1D channels. The most interesting possibilities arise in the presence of a magnetic field, which 
controls
interference, due to the Aharonov-Bohm (AB) effect. This effect manifests itself  in the bulk properties of  2D topological insulators \cite{Peng2010,Lin2017,Bardarson2010,Bardarson2013, KvonAB2015} as well as in the   helical edge state (HES) systems  and interferometers based on them \cite{Buttiker2012,Dolcini2011,KvonAB2015,Niyazov2018,Niyazov2020,Niyazov2021,Niyazov2021a}.

 \begin{figure}[b]
 \includegraphics[width=0.95\columnwidth]{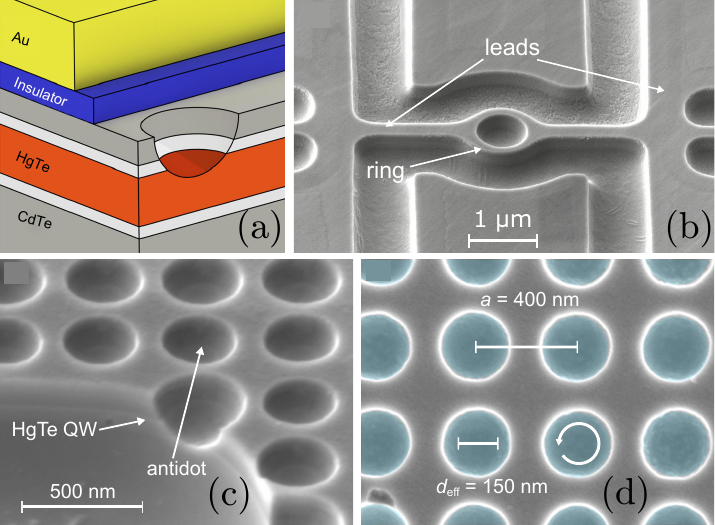}
\caption{\label{fig:antidots} 
 Various structures based on antidots  (etched holes) in the  topological insulator: (a) etched hole  in HgTe/CdTe quantum well,  (b) single etched hole and etched channel connected to leads, (c)  antidot lattice in  8 nm width HgTe quantum well, (d)  helical crystal formed by HES on the boundaries of holes; [panel (a) is adopted from \cite{Maier2017},  panels (b)-(d) are adopted from \cite{Ziegler2019}].}
\end{figure}
  
The interference   is usually suppressed  at high temperature, $T.$  However,  in the HES-based systems   this limitation   is sufficiently  soft,  particularly  due to progress in fabrication of high-quality  2D topological insulators.   
Specifically, for good quantization,  $T$  should be much smaller than the bulk gap of the topological insulator. 
The quantum spin Hall effect was first observed in structures based on HgTe/CdTe~\cite{confirmed} and InAs/GaSb~\cite{Knez2011},  where the bulk gap is rather small, less than 100 K. Substantially large values of the gap of  order of 500 K were observed  in WTe$_2$  ~\cite{Wu2018}; while a bulk gap of about 0.8 eV was demonstrated in bismuthene grown on a SiC (0001) substrate  ~\cite{Reis2017,Li2018} (see also recent discussion in Ref.~\cite{Stuhler2019}).    These experiments  indicate the possibility of topologically  protected  transport through helical channels even at room temperature.   One could also  naively expect that interference is damped when $T$ is larger than the level spacing.   As was  recently demonstrated  theoretically \cite{ Niyazov2018, Niyazov2020,Niyazov2021,Niyazov2021a}, this is not the case for AB HES-based  interferometers, where interference  survives even for the case $T\gg  2\pi v_{\rm F}/(L_1+L_2)$ where  $L_{1,2}$ are lengths  of the interferometer   shoulders and $v_{\rm F}$ is the Fermi velocity.  This implies that the interference effects  in the HES-based systems can be studied at  relatively high temperatures practically relevant for various applications.             

 \begin{figure}
\includegraphics[width=0.97\linewidth]{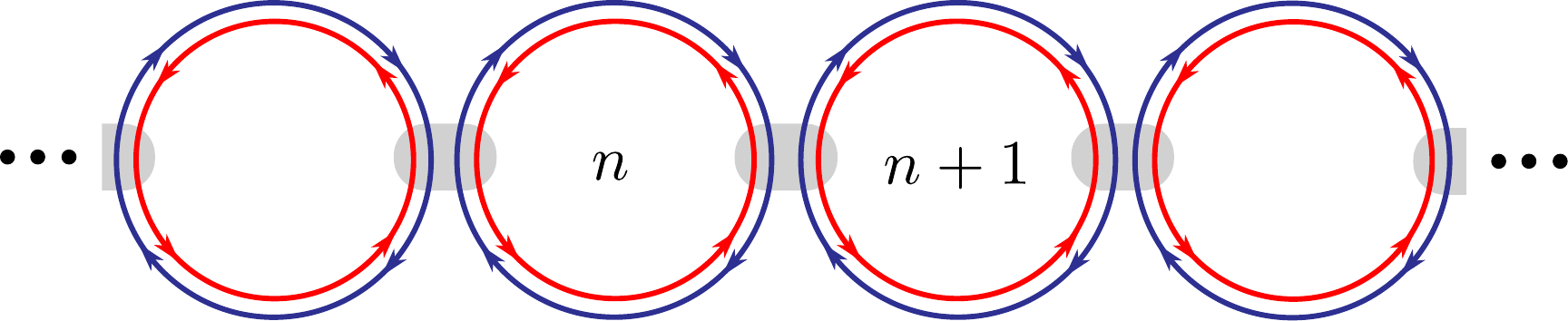} 
\caption{\label{fig:1Crystal}
Chain of helical rings with HES (shown by red and blue color)  coupled by tunneling or ballistic  contacts (shown by grey color). 
}
\end{figure}

  \begin{figure}
\includegraphics[width=0.99\linewidth]{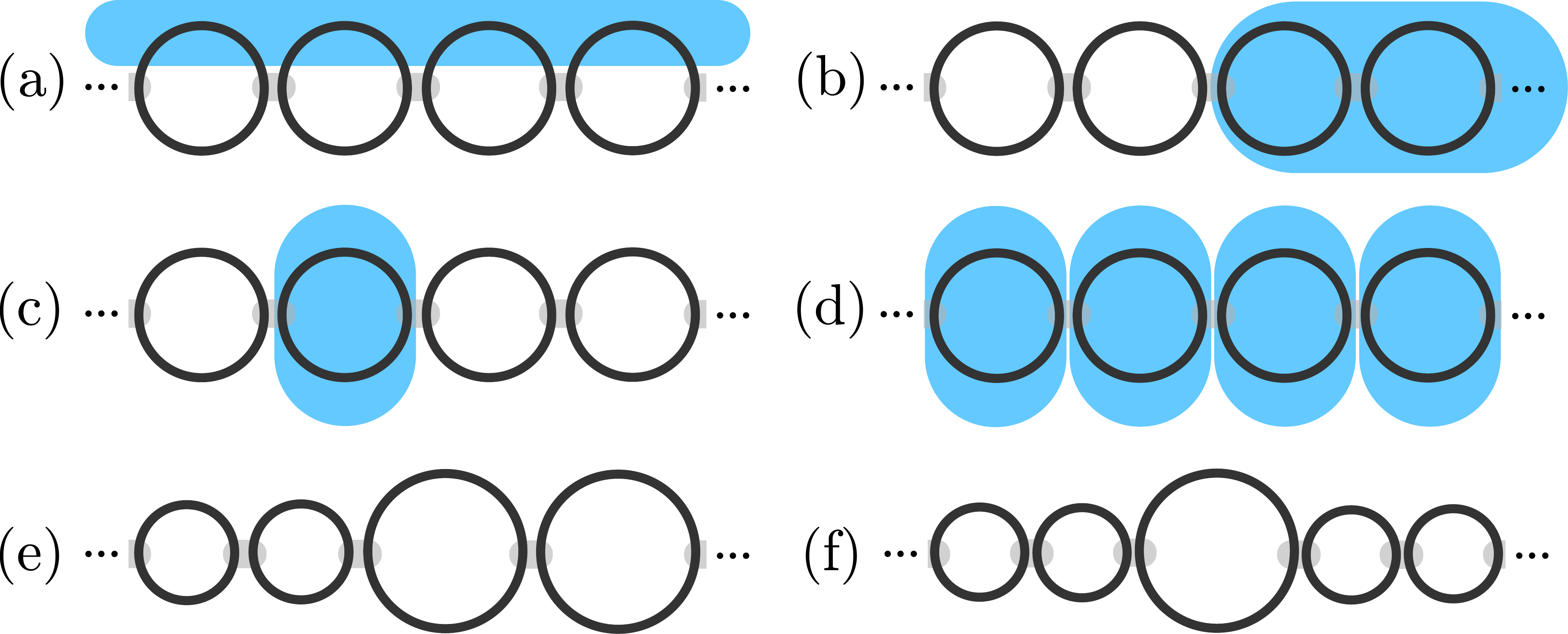}
\caption{\label{fig:nring}
(a-d) Various gate configurations (gates are shown by blue color)  allowing  to tune the crystal: (a)   control  of up-down symmetry, (b)   control  of right-left symmetry, (c)  voltage shift of  an individual ring, (d) back gate controlling Fermi level and tunneling transparency of the contacts between rings;  (e-d) defects in the crystal: (e) a  wall separating rings of different radii encompassing different fluxes,      (f) ``defect'' ring of large radius with   a higher flux. 
}
\end{figure}

\begin{figure}
 \includegraphics[width=0.7\columnwidth]{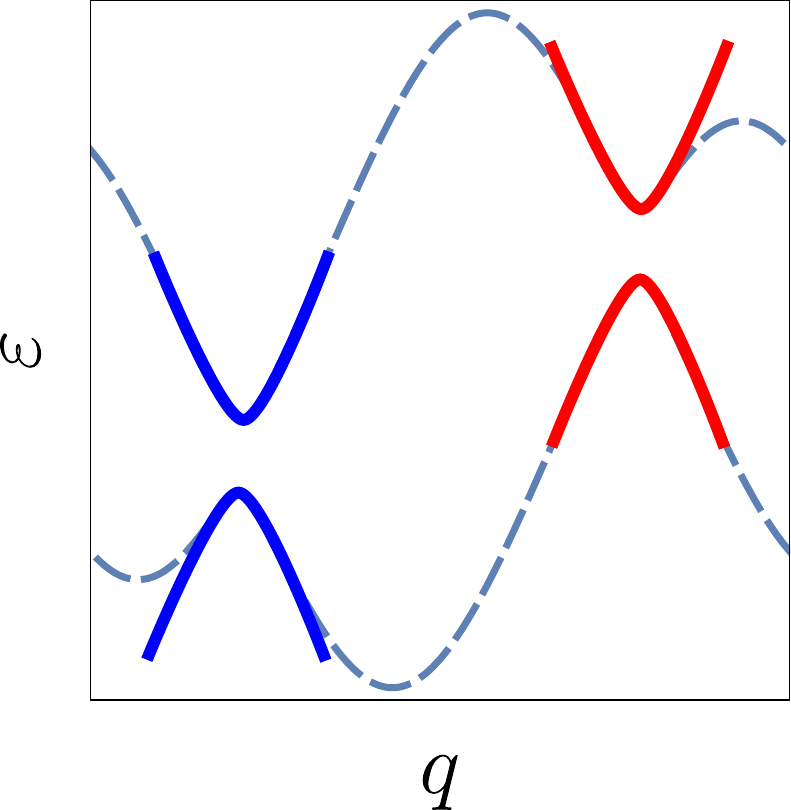}
\caption{\label{fig:dpintro}
  Schematical  dispersion of a band of the  helical crystal.   
Two massive Dirac cones are marked by blue and red color.}
\end{figure}

One of the promising  direction of  further research is the study of the HES-based metamaterials such as helical crystals.   First experimental realizations of such systems were  reported  recently \cite{Maier2017,Ziegler2019}. In these works,  the transport properties of an array of  holes chemically etched  in  the layer of   topological material (see Fig. \ref{fig:antidots}) were studied. 
Two different types of AB magnetosocillations with the periods $h/ec$ and $h/2ec$ were observed. 

The distance between holes in these experiments was sufficiently large, so the holes were not tunnel-connected.   On the other hand,  similar structures with  a shorter  distance  and, therefore, stronger   tunneling coupling  between   neighboring etched holes  could     demonstrate      tunneling transport.   
Moreover, tunneling barriers between holes can be controlled by  gate electrodes.  This way, by using 2D topological insulator with etched holes one can  fabricate  one- or two-dimensional  array of tunnel-coupled  HES,      i.e.  a  helical crystal tunable by gate electrodes.

In this work, we consider the simplest realization of the helical crystal, namely, 1D periodic array of closely placed identical etched holes. Specifically, we consider  transport through the superlattice formed by HES existing on the boundaries of these etched holes and coupled to each other by 
contacts, characterized by general scattering matrix,
see Fig.~\ref{fig:1Crystal}. The most important property of such a crystal is its tunability. The band structure  of the crystal can be  controlled by both gate electrodes and external  magnetic field.
Indeed, the external magnetic field  can change  flux through a hole thus  changing the  interference conditions.  An additional control is  achieved by using different types of  gate electrodes (see Fig.~\ref{fig:nring}).
Modern technology allows one to apply both back gate covering all system or part of the system and individual gates to holes thus creating ``defects'' in the crystal  structure or periodic modulation of the structure.
Importantly,  tunnel-coupling is not the only way to connect  HES belonging to different etched holes into a crystal. Effective coupling between holes  in the regime of bulk insulator  can be also realized  by fabricating  ballistic contacts between  holes  made from normal (non-topological)   materials.  Then, by using special configuration of the gate electrodes one could change transparency (scattering matrix) of such contacts.
 
We will demonstrate that the  variation of magnetic field allows one to create Dirac points (DPs) in the spectrum.  Such points appear at special  values of the field corresponding  to integer or  half-integer magnetic flux quantum through the hole. The  positions and  velocities of the Dirac cones   are tuned by gates   while deviation of the   flux from the special values leads to formation of gaps, i.e.  generation of the finite mass of the  Dirac fermions (see Fig.~\ref{fig:dpintro}).   With further deviation of the    dimensionless  flux   from the integer or half-integer values,   different DPs and corresponding massive Dirac cones can converge  forming a double DP. In vicinity of such point there are two  massive Dirac cones that disappear, when flux deviation  becomes higher than some value.  We also find that defects in the crystal (see Fig.~\ref{fig:nring} e,f) may lead to emergence  of topologically protected states of  the Volkov-Pankratov type~\cite{Volkov1985}. At some conditions specified below these  Volkov-Pankratov states  are double  degenerate.  However, the degeneracy can be lifted by changing magnetic field  or,   more importantly, in a purely electrical way,  by  changing potentials of gates.   This means formation of a topologically protected qubit which   can   be manipulated by gates.  
 
\section{Model}
Let us specify our model. 
We study an 1D array of antidots (etched holes)  on the surface of the 2D spin Hall insulator.  HES exist at the boundaries of the holes.  Electrons propagate along  boundaries of the holes thus  forming a chain of  identical helical rings of length $L$  connected in series (see Fig.~\ref{fig:1Crystal}). Each ring contains clockwise- and counterclockwise-moving electrons with energy,  $E,$  and momentum, $k= E/v_{\rm F},$   having  spins opposite at each point.  We assume the presence of the  external magnetic field and piercing  each ring by flux $\Phi.$    
 
We assume that $X$-junction between two neighboring rings with four scattering channels     (see Fig.~\ref{fig:S-matrix}) is described by the scattering matrix $\hat S$  preserving symmetry with respect to time  inversion \cite{Teo2009,Aristov2016}:   
\begin{figure}
\includegraphics[width=0.43\columnwidth]{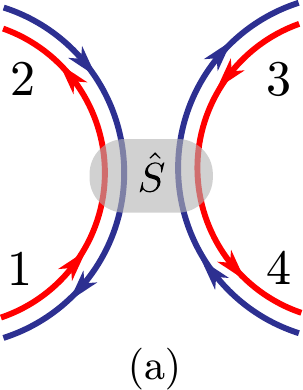} \hfill
\includegraphics[width=0.47\columnwidth]{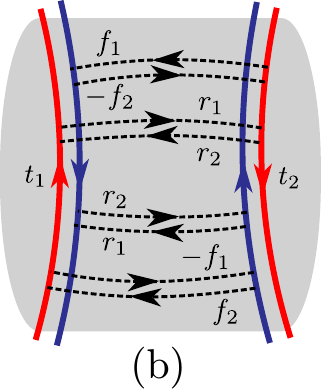}
\includegraphics[width=0.42\columnwidth]{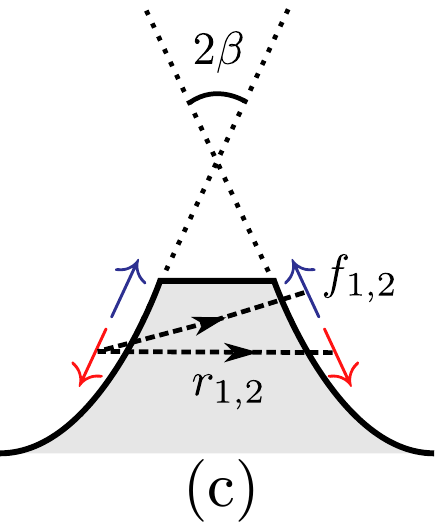}
\caption{\label{fig:S-matrix}
(a) Junction between two neighboring rings, (b) Amplitudes, $t_i,~r_i,~f_i$  ($i=1,2$),  of different  scattering processes are described by elements of  $\hat S-$matrix   Eq.~\eqref{eq:smat}, (c) For   non-zero angle $\beta,$ the spin-flip processes are allowed with the amplitudes $f_i \propto \sin\beta.$   }
\label{rtf}
\end{figure}

\be
C_\alpha^\prime= S_{\alpha \beta} C_\beta, \quad \alpha,\beta=1,2,3,4;
\ee 
\aleq{ \label{eq:smat}
\hat S&=\begin{pmatrix}
0 & t_1 & f_1 & r_1 \\
t_1 & 0 & r_2 & f_2 \\
-f_1 & r_2 & 0 &   t_2 \\
r_1& -f_2 &   t_2 & 0  \\
\end{pmatrix} \,, \\
t_j&=t e^{i \tau_j} \,, \quad
f_j=f e^{i \phi_j} \,, \quad
r_j= i\, r e^{i \gamma_j} \,.\\
}
Here $C_\alpha^\prime$  and $C_\beta$ are amplitudes of  out-going and incoming waves,  respectively,  and
$t$, $f$, $r$ are real-valued. 
The elements of the scattering matrix obey  
\aleq{
&t^2+r^2+f^2=1 \,, \\
&\tau_1+\tau_2=\phi_1+\phi_2=\gamma_1+\gamma_2 \,,
\label{eq:unitarity}
}
due to unitarity. One can show that the spectrum of the crystal depends only on combinations $\tau_1+\tau_2$ and $\phi_1+\phi_2,$ so that  without loss of generality and using \eqref{eq:unitarity} one can choose
\be
\tau_1=\tau_2=\phi_1=\phi_2=\frac{\gamma_1+\gamma_2}{2}.
\label{unitarity}
\ee
 
Next, we introduce the transfer matrix $\hat T$ describing a transition from $n-$th to $(n+1)-$th cell of the helical crystal (see Appendix \ref{app:transfer}), 
\be 
\hat T =  \hat P_0 \hat T_0 
\label{T}
\ee 
where  $T_0$ and $P_0$ describe transitions through the contact and the ring, respectively.

The  matrix of contact,  $T_0$,
corresponds to the  matrix $\hat S$  and  expresses the wave function amplitudes in the $(n+1)-$th ring (right after the contact)  via the amplitudes in $n-$th  ring (right before the contact). It
can be written as follows   
\aleq{
\hat T_0 &= \hat D \hat T_{\text{r}}\hat D \, ,\\
\hat T_{\text{r}} &=
\frac{1}{1-t^2}\begin{pmatrix}
ir & f t &  {-}f & -irt \\
-f t& {-}ir & i r t &   f \\
 {-}f & - i r t & ir  & f t \\
i r t  &  f  & -f  t & {-}ir  \\
\end{pmatrix}\,,\\
\hat D&=\text{diag} [ e^{i \gamma_2/2},e^{-i \gamma_2/2},e^{i \gamma_1/2},e^{-i \gamma_1/2} ] \,.\\
\label{T0}
} 
The matrix
\aleq{
\hat P_0=\text{diag} [e^{i \varphi_b},e^{-i \varphi_a},e^{i \varphi_a},e^{-i \varphi_b}]\, ,
\label{def:Pmat}
}
contains phases acquired inside a ring by moving from the left contact to the   right one:
\be
\varphi_a= \frac{\varepsilon}{2} - \pi \phi, \quad
\varphi_b= \frac{\varepsilon}{2} + \pi \phi,
\label{phases}
\ee
where we introduced dimensionless energy and dimensionless flux:
\be
\varepsilon= \frac{E L}{v_{\rm F} }= k L, \qquad   \phi=\frac{\Phi}{\Phi_0},
\label{dimen-E-Phi}
\ee
where  $\Phi_0$ is   the flux quantum.    

In Eq.~\eqref{phases} we assumed that etched holes are ring-shaped,   contacts are placed symmetrically,  and energies of the electrons in the up and down parts of the ring are the same. However, by using specific gate configurations (see Fig.~\ref{fig:nring}a) one can change energies, which  would lead to appearance of two different dimensionless energies,  $\varepsilon^{\rm up}=k^{\rm up}L $ and $\varepsilon^{\rm down}=k^{\rm down}L $ (see Appendix \ref{app:transfer}).           
 
The amplitudes $t$, $r,$  and $f$  can be parametrized as follows: 
\aleq{
t&=\cos \alpha \,, \\
r&= \sin \alpha \cos \beta  \,,\\
f&=\sin \alpha \sin \beta \,.
\label{paramTRF}
}

Processes corresponding to scattering amplitudes    are illustrated in  Fig.~\ref{fig:S-matrix}. Amplitudes   $r$ and $f$ describe hopping between rings into the states with the same and opposite spin projections, respectively.  Physically,  $\alpha$ is related to the total  probability of such hopping: $r^2+f^2=\sin^2 \alpha.$  
The angle  $2\beta$ is the angle between the spin quantization axes in different helical states~\cite{Aristov2017}, and  elements  $r$ and  $f$ are proportional to    overlapping of spinors in the neighboring helical edges.  
For tunneling contact  ($\alpha \ll 1$),   $\alpha$  is  mainly controlled  by tunneling  distance.  The case of ballistic (or metallic) contact, when rings are strongly coupled,  corresponds to $\alpha \approx \pi/2.$  Parameter $\beta$ controls spin-flip  processes. Indeed, for $\beta=0,$ we get $f=0.$         

\section{Dispersion of 1D helical crystal}
Next, we derive and analyze  the dispersion relation of  1D helical crystal  shown in Fig.~\ref{fig:1Crystal}, i.e. the  dependence $\varepsilon=\varepsilon(q)$  of the  dimensionless energy  $\varepsilon$    on  the   dimensionless quasi-momentum $q$ measured in units of $1/L.$

\subsection{General dispersion equation}
We look for those four-component vectors, $\psi_j$, referring to $j$th ring, which upon the action of transfer matrix return to their form up to a phase factor, that is $\psi_{j+1} = \hat T \psi_j= e^{iq} \psi_j$.
Hence the band structure of the   crystal 
 is found from   the  following  equation: 
\aleq{\label{eq:dispeq}
\det [\hat 1 e^{iq} -\hat T]=0 \,,
}
 where $\hat 1$ is $4\times 4$ unit matrix. 
The resulting dispersion relation, $\varepsilon(q)$, is quite complicated
and can be found  in implicit form by using Eqs.~\eqref{T},\eqref{T0} and \eqref{def:Pmat}:  

\aleq{
&\left(\cos \beta  \sin \left(  {\varepsilon }/2+\pi  \phi \right)+\sin \alpha  \cos (q- v/2)\right) \\
&\times \left(\cos \beta  \sin \left( {\varepsilon }/2-\pi  \phi \right) +\sin \alpha  \cos (q+ v/2 )\right)\\
&+\tfrac{1}{2} \sin ^2\beta  \left(\cos ^2\alpha  \cos 2 \pi  \phi -\sin ^2\alpha  \cos v -\cos \varepsilon \right)=0.    
\label{disp-general}
} 
  Here,
\be 
\varepsilon= kL+ \gamma_1+ \gamma_2,\quad  v= \gamma_2-\gamma_1 
\label{gamma1gamma2}.
\ee
 We note a following general property: if $q$, $\varepsilon$ satisfy  Eq.~\eqref{disp-general}, then   $q\to q + \pi$, $\varepsilon\to \varepsilon+ 2 \pi$ also satisfy this equation.
 
 As seen from Eqs.~\eqref{disp-general} and \eqref{gamma1gamma2},  the sum  $\gamma_1+\gamma_2$  affects only the total  energy shift, so that without loss of generality we can set  $\gamma_1+\gamma_2=0.$
 In this case we have 
\be
\gamma_2=-\gamma_1=\frac{v}{2}.
\ee
 In order to understand the physical meaning of the phase $v,$ one can consider different energies at the upper and lower shoulders of the ring. Then, $\hat P_0$ is given by  a more general equation Eq.~\eqref{def:Pmat-general}. One can check that Eq.~\eqref{disp-general} still holds with 
\aleq{
\varepsilon &= {(k^{\rm up} +k^{\rm down})L}/{2}+ \gamma_1 +\gamma_2 , \\ 
 v & =   {(k^{\rm up} -k^{\rm down})L}/{2}+ \gamma_2 -\gamma_1. }
 Both  $k^{\rm up}$ and $ k^{\rm down}$ depend on the electron energy, $E$, and can be also controlled by the gate voltages,  so that   
 $k^{\rm up} -k^{\rm down} \propto  U^{\rm up} -U^{\rm down},$ 
 where    $U^{\rm up, down}$ are the gate  voltages applied to the upper and lower shoulders  of the interferometer  [see Fig.~\ref{fig:nring}(a)].  Hence, Eq.~\eqref{disp-general} is equally applicable to  the system with  complicated gate configurations. 
 The expression for $v$ shows that the difference of  gate voltages is equivalent to difference of $\gamma_1$ and $\gamma_2.$ It means that the parameter $v$ controls the up-down asymmetry of the system, arising either due to asymmetry of the contact, $\gamma_2-\gamma_1$, or due to asymmetric gate voltage configuration  ($k^{\rm up} -k^{\rm down} \propto  U^{\rm up} -U^{\rm down}$).       For simplicity, one can assume that $k^{\rm up} =k^{\rm down}=k=E/v_{\rm F}.$   
 Then, Eq.~\eqref{disp-general} gives dependence of the dimensionless energy $\varepsilon$ on $q,$ while the dependence on $v$ is equivalent on the dependence on the gate voltage difference  $U^{\rm up} -U^{\rm down}.$

 \begin{figure*}
\includegraphics[width=0.87\linewidth]{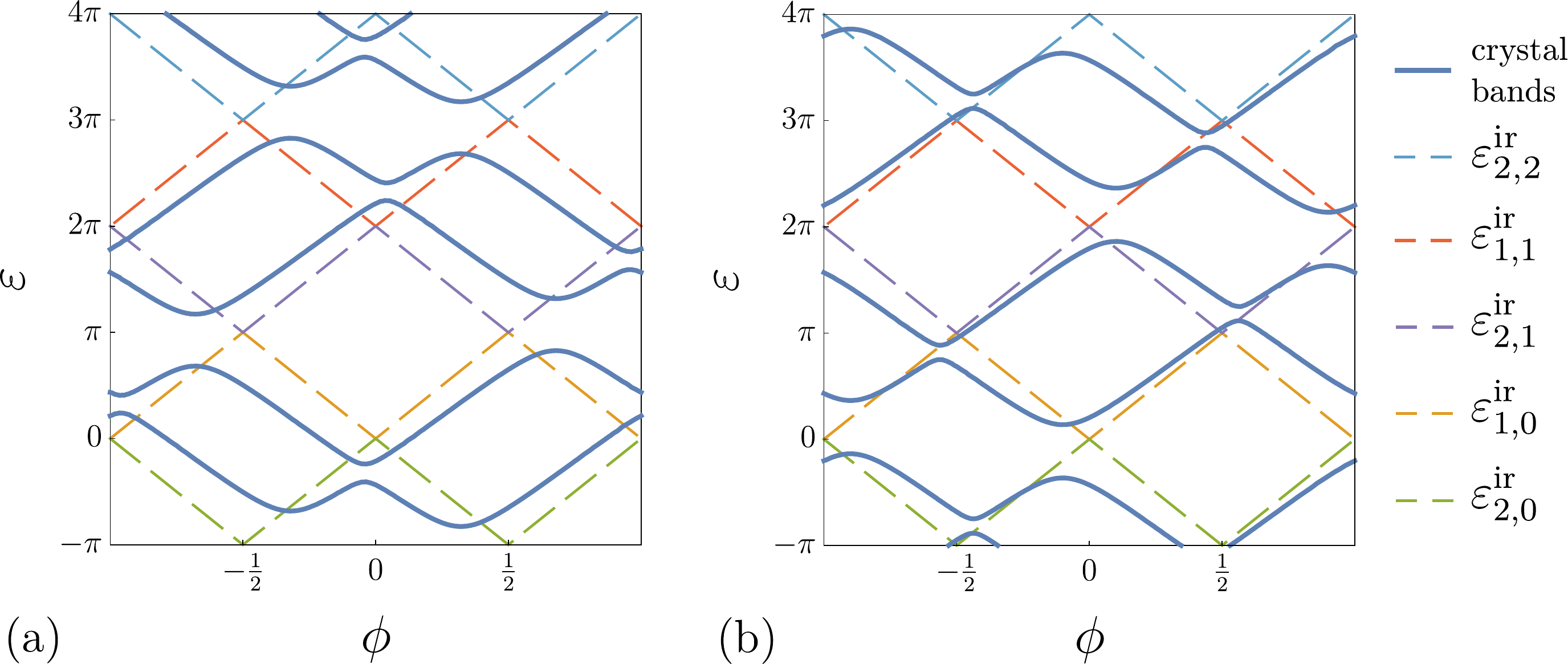}  
\caption{\label{fig:Ephi}
Dependence of the dimensionless energy  in several neighboring  bands  of  helical crystal    on the magnetic flux for  fixed $q$  at  $\alpha = \pi/4$,  $v=\pi/3$ and  $\beta=0.6$; (a): $q=\pi/10,$ (b): $q=2\pi/5.$ Dashed curves correspond to  $\alpha=0$ and  are described by Eq.~\eqref{levels0}.    }
\end{figure*}

 \begin{figure*}
\includegraphics[width=0.95\linewidth]{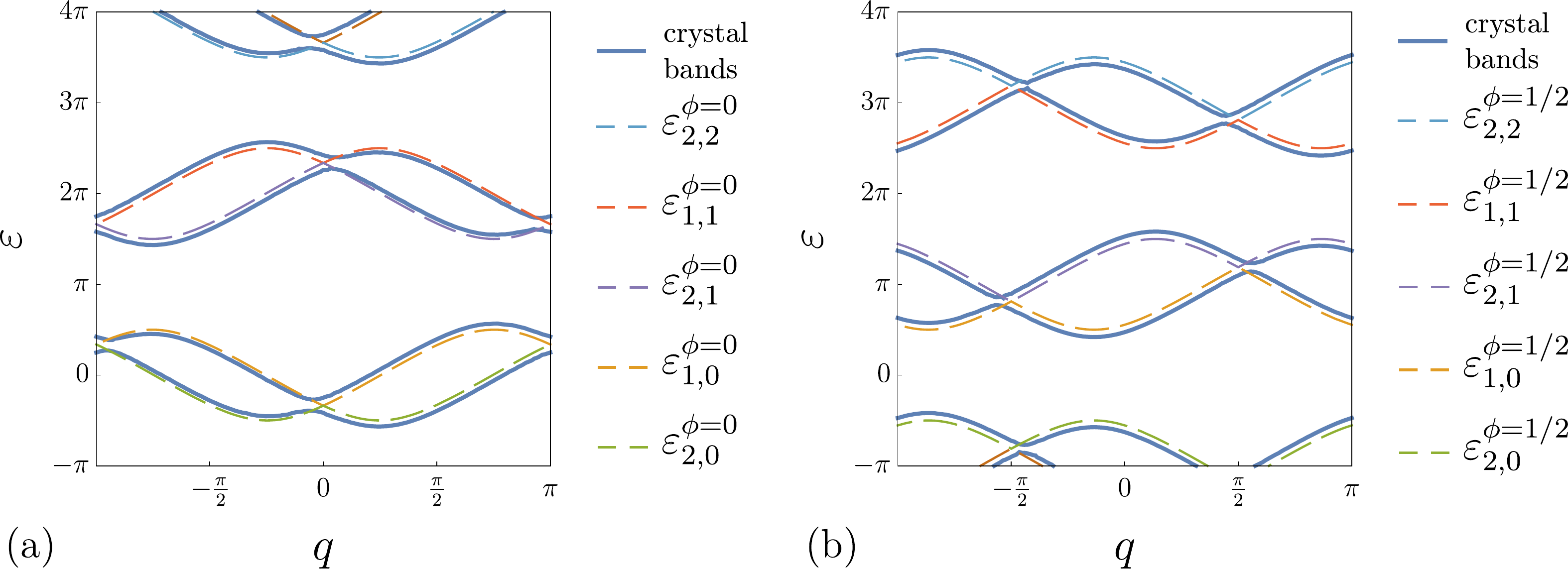} 
\caption{\label{fig:DPphi}
Several lowest energy bands  of the helical crystal for $\alpha = \pi/4$, $v=\pi/3,$ and $\beta=0.6$; (a): $\phi=0.05$ [dashed curves corresponds to $\phi=0$ and are described by Eq.~\eqref{eq:energy_phi_0}];  (b): $\phi=0.45$ [dashed curves corresponds to $\phi=1/2$ and are described by Eq.~\eqref{bands-phi-12}].  
}
\end{figure*}
 
Next, we discuss different limiting cases covered by  Eq.~\eqref{disp-general} and corresponding to different physical situations.

\subsection{Independent rings, $\alpha=0$}  
In the absence of the tunneling coupling, i.e.\ for $\alpha=0,$   the rings are isolated.  Hence, all bands of the crystal shrink  into  the following quantum levels for clockwise- and counterclockwise- moving electrons, respectively:
\be
\varepsilon_{n}^{  \rm  +}= 2\pi (n+ \phi),\quad \varepsilon_{n}^{  \rm  -}= 2\pi (n- \phi),
\label{levels}
\ee
with integer  $n$ [here we do not write the  shift of the total dimensionless energy $\gamma_1+\gamma_2=0;$ see discussion after Eq.~\eqref{disp-general}].   
As seen from Eqs.~\eqref{levels}, the most interesting effects are expected for  $\phi=0$ and  $\phi=1/2,$ when levels are doubly degenerate. 

For small but finite $\alpha$ ($\alpha \ll 1 $), the energy levels \eqref{levels}    
evolve into narrow bands,  with the width proportional to $\alpha.$   
In order to study dependence of bands on $\phi,$ it is convenient to  parametrize the energy levels at $\alpha=0$ in the following way  
 \be
 \begin{aligned}
&\varepsilon_{1,n}^{\rm ir}= 2\pi n +  \arccos (\cos 2\pi \phi ) \,
\\
&
\varepsilon_{2,n}^{\rm ir}= 2\pi n - \arccos (\cos 2\pi \phi )\,
\end{aligned}
\label{levels0}
\ee
where label ``ir'' stands for ``independent rings''.
This parametrization  is different from Eq.~\eqref{levels}. The key advantage of Eq.~\eqref{levels0}  is the possibility to describe the
 avoidance of level crossing which happens at $\alpha \neq 0$ for $\phi \approx 0$ and $\phi \approx 1/2$  as we demonstrate below. 
We notice, that Eq~ \eqref{levels0}, unlike Eq~\eqref{levels},  is periodic in $\phi$ and has the property   
$\varepsilon_{1,n}^{\rm ir} > \varepsilon_{2,n}^{\rm ir}$ 
for any $n$ and $\phi$. 

\subsection{
Independent shoulders}

The case $\alpha=\pi/2$, $\beta=0$ corresponds to the disconnected upper and lower shoulders of the array. As follows from  Eq.~\eqref{disp-general}, there are four branches of  dispersion  in this case, which are given by 
\aleq{
\varepsilon_{1,n}^{\rm is} &= 2 q +2\pi \phi + v + \pi (4n-1) \,, \\
\varepsilon_{3,n}^{\rm is} &= -2 q - 2\pi \phi + v + \pi (4n-1) \,, \\
\varepsilon_{2,n}^{\rm is} &= 2 q -2\pi \phi - v + \pi (4n-1) \,, \\ 
\varepsilon_{4,n}^{\rm is} &= -2 q +2\pi \phi - v + \pi (4n-1)\,,  
\label{pi2beta0}
}
where label ``is'' stands for ``independent shoulders''.
The  integer index $n$ corresponds to our consideration of reduced Brillouin zone scheme,   $q\in (-\pi,\pi)$. 
For disconnected shoulders, however, the consideration of the extended Brillouin zone scheme, $q\in (-\infty,\infty)$, is equally suitable. In the latter case we should  set $n=0$ in right-hand side of Eq.~\eqref{pi2beta0} to avoid double counting.  

The meaning of four branches is easily understood.
They correspond to spin-conserving  chiral  motion.
The branches $1$ and $3$ correspond to right- and left-moving electrons, respectively, in the upper shoulder.   The branches $2$ and $4$ refer to right- and left-moving electrons, respectively, in the lower shoulder.

\subsection{Spin conserving motion}
Let us now generalize the case considered in the previous subsection by assuming that  $\alpha \neq \pi/2$ but still $\beta=0.$  This case corresponds to  hopping between rings without spin flips. From Eq.~\eqref{disp-general} we find that there are several solutions found from 
\aleq{
&  \sin \left(  {\varepsilon }/2+\pi  \phi \right)+\sin \alpha  \cos (q- v/2)  = 0 \,,  \\
&   \sin \left( {\varepsilon }/2-\pi  \phi \right) +\sin \alpha  \cos (q+ v/2 ) =0.     
}
These equations yield four branches of dispersion: 
\aleq{
\varepsilon_{1,n}^{\rm sc} &= -2 \arcsin[\sin\alpha\cos(q+v/2)] +2\pi \phi  + \pi (4n) \,, \\
\varepsilon_{3,n}^{\rm sc}  &=  2 \arcsin[\sin\alpha\cos(q-v/2)] -2\pi \phi  + \pi (4n-2)    \,, \\
\varepsilon_{2,n}^{\rm sc}  &= -2 \arcsin[\sin\alpha\cos(q-v/2)] -2\pi \phi  + \pi (4n)  \,, \\ 
\varepsilon_{4,n}^{\rm sc}  &= 2 \arcsin[\sin\alpha\cos(q+v/2)] +2\pi \phi  + \pi (4n-2) \,, 
\label{sc-branches}
}
where label ``sc'' stands for ``spin conserving''. The choice of branches  here is done in such a way that $\varepsilon_{j,n}^{\rm sc}$ coincides with $\varepsilon_{j,n}^{\rm is}$ [see Eq.~  \eqref{pi2beta0}] for $\alpha = \pi/2$  and $(q \pm v/2)\in(0,\pi)$. 
It is worth noting that for arbitrary $q$ and $v$ the equations \eqref{pi2beta0} and \eqref{sc-branches} give different parametrization of the spectrum for $\alpha=\pi/2.$  Equation \eqref{sc-branches} is more convenient because it allows to describe the avoidance  of  level crossing, happening for arbitrary small deviation of $\alpha$ from the value $\pi/2$          [in   a  full analogy  with the difference between  Eqs.~\eqref{levels} and \eqref{levels0}].

\subsection{Perfect spin-flip chiral motion. } \label{sec:perfectspinflip}
For $\alpha=\pi/2$ and  $\beta=\pi/2,$ the  spectrum does not depend on the flux and  is given by 
\aleq{
\varepsilon_{1,n}^{\rm sf} &= 2 q  + 2\pi n \,, \\
\varepsilon_{2,n}^{\rm sf} &= -2 q  + 2\pi n \,,  
\label{pi2pi2}
}
where label ``sf'' stands for ``spin flip.''
In this case,    the chirality is conserved, but  the  electron jumps between shoulders with flipping  the  spin  direction   after transmission of any contact. Such unusual transmission can be induced by the  effects of interaction (see Section~\ref{interaction}).

\subsection{Special values of magnetic flux.}
The spectrum dramatically  simplifies for integer and half-integer values of the dimensionless magnetic flux.   Most interestingly the Dirac points appear in all bands.  These   points   appear by pairs (one pair  per Brillouin zone for any band). They  are doubly degenerate for $v=0.$          

Before we turn to the analysis of these special values of the flux, we notice that  Eq.~\eqref{disp-general} is invariant under the replacement 
\be
\phi \to \phi +1,\quad q \to  q  +\pi. 
\ee
It follows that we can restrict our consideration  to the cases  $\phi=0$ and $\phi=1/2.$   Let us find  the bands of the crystal exactly  at these values of flux. 
 
\subsubsection{$\phi=0.$}
 
 In this case,  Eq.~\eqref{disp-general} is reduced to 
\be
\label{eq:limit_phi-0}
 \sin [\varepsilon/2] =-\sin \alpha \cos (q \pm q_0), 
\ee
where $q_0$ ($0<q_{0}<\pi$) obeys  $ \cos q_0= \cos \beta \cos(v/2). $  
Solutions of Eq.~\eqref{eq:limit_phi-0}  can be written as  
\aleq{
\label{eq:energy_phi_0}
 \varepsilon^{ \phi=0} _{1,n} = & 2 \pi n +  2 \arcsin \left[  \sin \alpha ((-1)^{n+1}\cos q \cos q_0 \right. \\
& \left.
+   |\sin q| \sin q_0) \right] \, ,
\\
 \varepsilon^{ \phi=0} _{2,n} = & 2 \pi n +  2 \arcsin \left[  \sin \alpha ((-1)^{n+1}\cos q \cos q_0 \right. \\
& \left.
-   |\sin q| \sin q_0) \right] \, ,
}

As one can  see from these equations, there are DPs in the spectrum corresponding to the condition $\varepsilon^{\rm \phi=0}_{1,n}=\varepsilon^{\rm \phi=0}_{2,n}.$  This condition is satisfied for 
\be
q_{\rm DP}= \pi m \, . 
\label{qDP-0}
\ee
with integer $m$ (for the first Brillouin zone  we keep $m=0,1$ only). 
Several lowest bands are  depicted in the left panel of    Fig.~\ref{fig:DPphi}.
  
\subsubsection{$\phi=1/2.$}
 
Substituting   $\phi=1/2$ into Eq.~\eqref{disp-general} we get
\be
\cos [\varepsilon/2] =-\sin \alpha \sin (q \pm q_1), 
\ee
where $\cos q_1= \cos \beta \sin (v/2)$ and $0<q_1<\pi.$ 
 
The energy bands found from this equation are given by
\aleq{ \label{bands-phi-12}
  \varepsilon^{\phi=1/2}_{1,n} & =  2 \pi n + 2 \arccos \left[ -  \sin \alpha (  (-1)^n \sin q \cos q_1 
\right. \\
& \left.
-   |\cos q| \sin q_1) \right] \, ,\\
\varepsilon^{\phi=1/2}_{2,n+1} & =  2 \pi n + 2 \arccos \left[ - \sin \alpha (  (-1)^n \sin q \cos q_1 
\right. \\
& \left.
+   |\cos q| \sin q_1) \right] \, .\\
}
There are Diracs points in the spectrum corresponding to the condition $\varepsilon^{\phi=1/2}_{1, n}=\varepsilon^{\phi=1/2}_{2, n+1}.$  Positions of these points are given by
\be
q_{\rm DP}=  \frac{\pi}{2}+\pi m \, , %
\label{qDP-12}
\ee
with integer $m$  ($m=0,-1$ for the first Brillouin zone).

Hence,   at special  values of the flux, $\phi=0$ and $\phi=1/2,$ 
the spectrum demonstrates the existence of Dirac cones.  These  cones are clearly seen in Fig.~\ref{fig:DPphi}. 
Importantly, when the flux deviates  from the special values, the DPs move away from the  values of $q$ given by  Eqs.~\eqref{qDP-0} and  \eqref{qDP-12} and a finite gap appears in the Dirac cones, so that these cones become massive. These points can  also converge with changing the flux forming double DPs and then disappear at certain value of the flux.

We discuss the massive Dirac cones and double DPs in more detail in subsequent Sections~\ref{sec:Dirac-points}, \ref{main-DDP}, and App. \ref{DDP}.

\subsection{Massive Dirac cones 
} \label{sec:Dirac-points}

The dependence on  $\phi$ for $\alpha \neq 0$ and fixed $q$  is shown in the Fig.~\ref{fig:Ephi}  by thick lines for small value of $q$ (left panel)  and for $q$ close to $\pi$ (right panel). As seen, there is level anticrossing at integer and half-integer values of flux.
 If we now fix flux values close to these special values, and change $q$, then  we get spectrum of the crystal with the DPs appearing at values of $q$ which are close to integer values of $q/\pi$  for  integer  $\phi$ and half-integer  values of  $q/\pi$  for  half-integer values of   $\phi$ as shown in Fig.~\ref{fig:DPphi}.

 Let us first discuss the spectrum for the case of nearly integer value of flux.
The position of DPs can be represented as :  
\aleq{
\varepsilon_\text{DP}&=2 \pi n - 2(-1)^{n+m+p}  \arcsin [\sin \alpha \cos \beta \cos (v/2)  ]\,, \\
\phi_\text{DP}&=p\,, \quad 
q_\text{DP} = \pi m\,,
 \label{DP:pos}
}
here $m = (-1,\,0),$ and $n,$ $p$ are integer.  
The spectrum close to all DPs can be written in the following form  
  \be
  \delta \varepsilon =  \pm
  \sqrt{(A \delta q + B \delta    \phi)^2 + C^2  (\delta  \phi) ^2   }.
  \label{spect}
  \ee
 where, $\delta \varepsilon,$   $\delta q,$ and $\delta\phi$  are counted from  their values exactly at DPs, Eq.\eqref{DP:pos}.
  
 As follows from this equation,  the gap in the spectrum is given by
 \be
 \Delta= 2 C |\delta \phi|,
 \ee
  The coefficients $B$ and $C$ are the same for all DPs with $\phi$ close to integer values,  while $A$ has different sign and the same absolute value for different DPs  : 
  \begin{align}
 \nonumber
  A &= 2~(-1)^{n+m+p}   \sin \alpha~ \sqrt{\frac{1-[\cos(v/2) \cos \beta]^2}{1- (\cos(v/2) \cos \beta \sin \alpha)^2 }}, 
 \\
  \label{ABC}
  B&=\frac{ 2\pi \cos\beta \sin(v/2) }{\sqrt{1- [\cos(v/2) \cos \beta ]^2}},
 \\
\nonumber
 C&= \frac{ 2 \pi \cos\alpha \sin \beta }{\sqrt{[1- [\cos(v/2) \cos \beta ]^2][1- [\cos(v/2) \cos \beta \sin \alpha]^2]}}.
 \end{align}
  We notice that $A\to 0$ for $\alpha \to 0,$ thus reflecting the existence of    flat 
 bands in this limit [see Eqs.~\eqref{levels}  and  \eqref{levels0}].   For $\beta=0$ the chiral channels are decoupled.  In this case,  $C=0$  and 
 \be
 \delta \varepsilon = |A \delta q+ B \delta \phi|.
 \ee
 Hence, in this special case we have a gapless spectrum for any $\delta \phi.$ 
 
The spectrum for the case of nearly half-integer values of flux $\phi$ is characterized by 
a different set of DPs, now written as 
 \aleq{
\varepsilon_\text{DP}&=2 \pi (n+\tfrac 12) + 2(-1)^{n+m+p}  \arcsin [\sin \alpha \cos \beta \sin (v/2)  ]\,, \\
\phi_\text{DP} &=m+\tfrac 12\,, \quad
q_\text{DP}= \pi p + \tfrac \pi2\,.
 }
Remarkably, Eqs.\ \eqref{spect}, \eqref{ABC} are still valid,  with the replacement     
 $v \to v+\pi.$

\subsection{Converging of the  Dirac points }\label{main-DDP}
\begin{figure*}
\begin{minipage}{0.33\linewidth}
\includegraphics[width=1\linewidth]{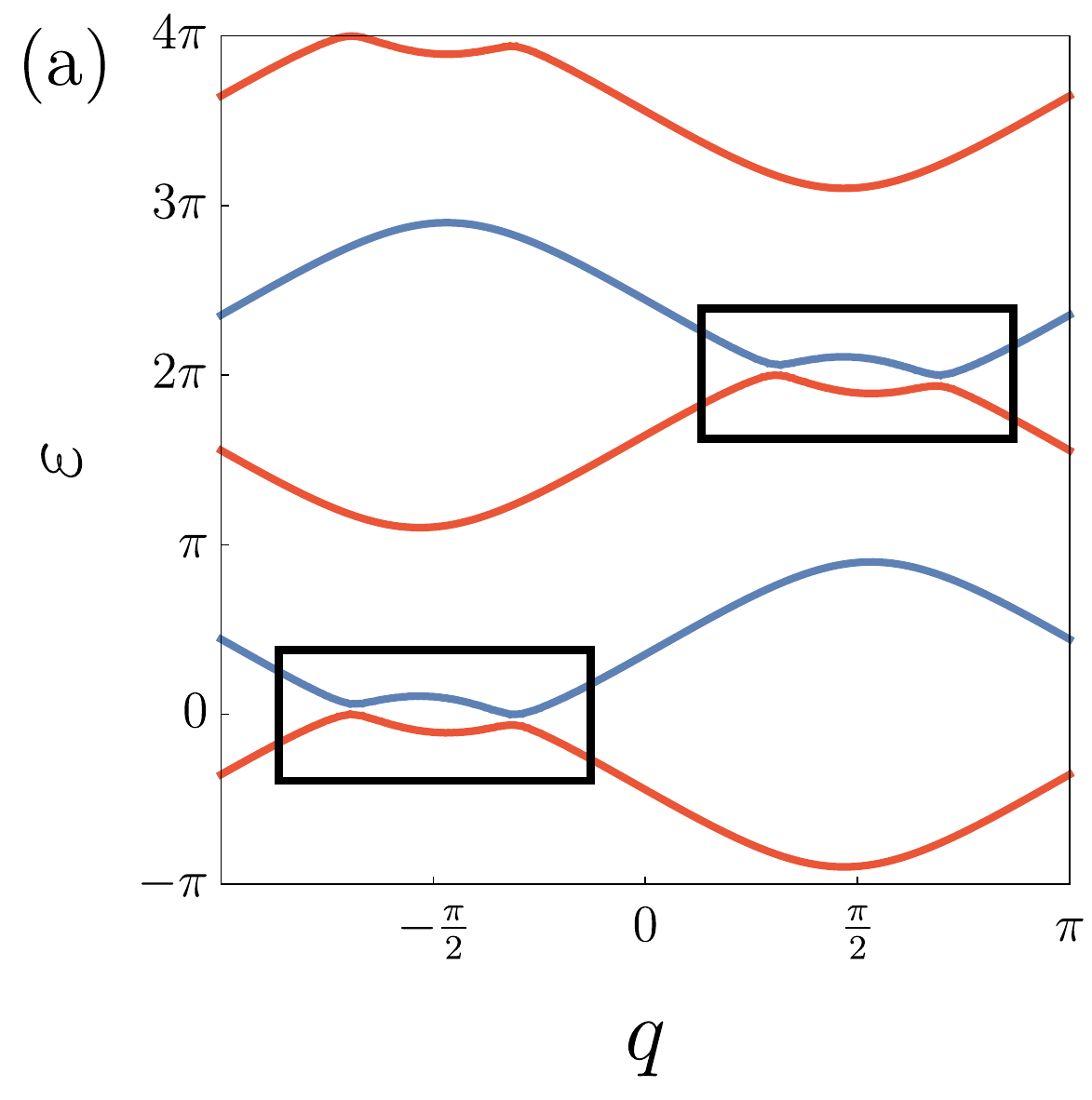} 
\end{minipage} 
\hfill
\begin{minipage}{0.29\linewidth}
\includegraphics[width=1\linewidth]{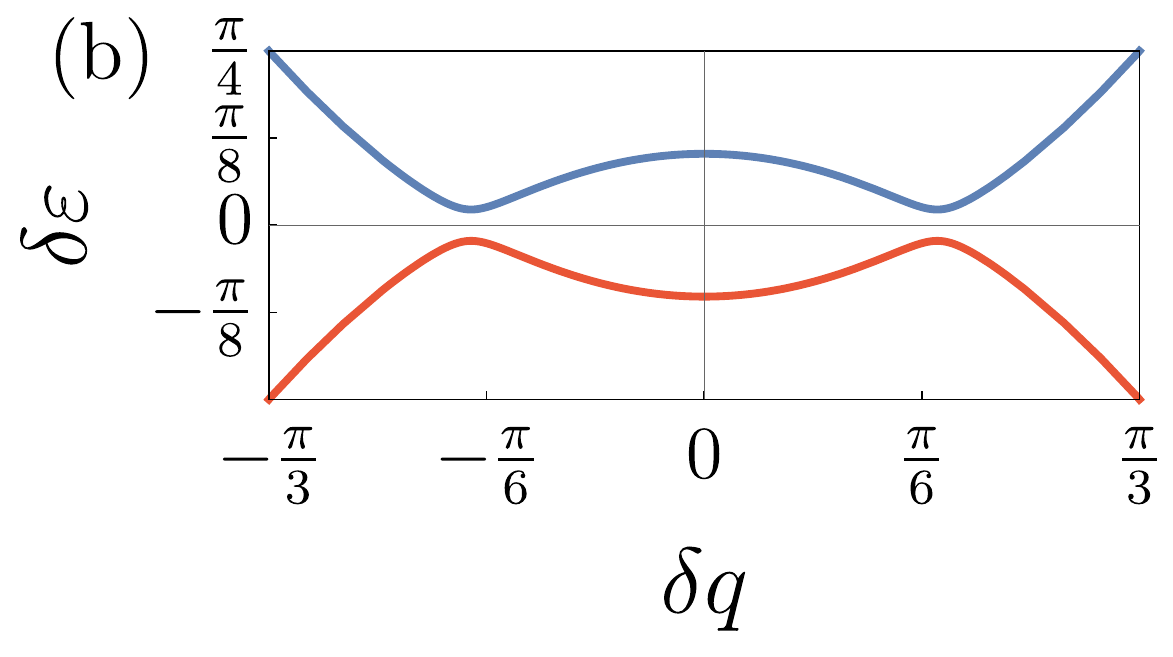} \vfill
\includegraphics[width=1\linewidth]{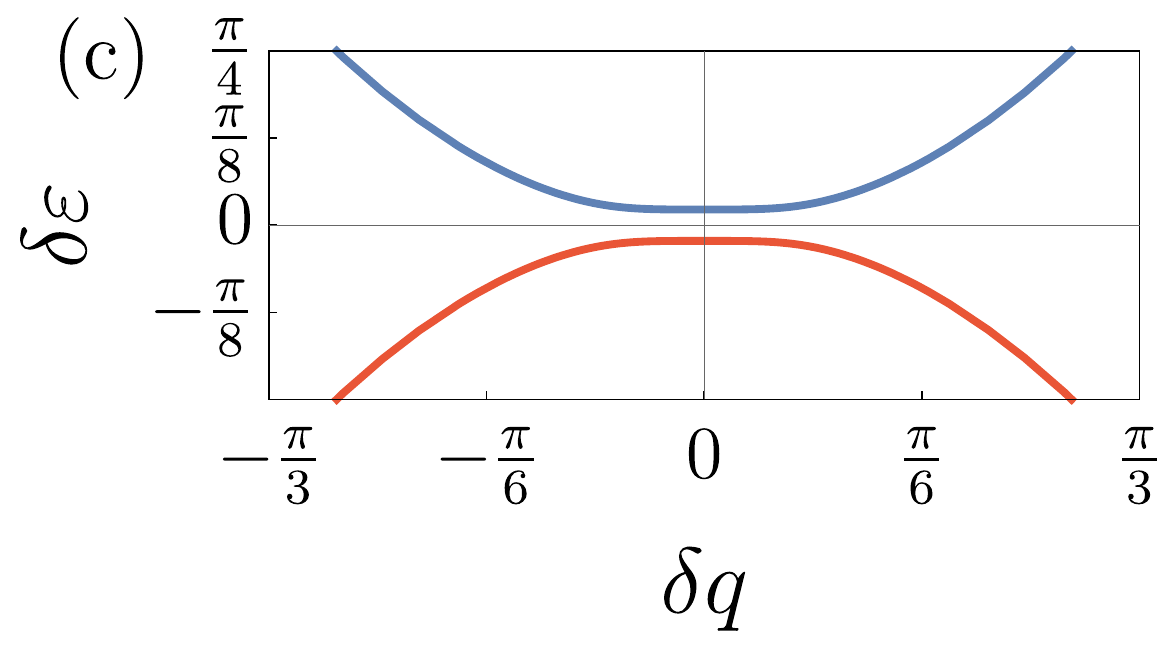} 
\end{minipage} 
\hfill
\begin{minipage}{0.29\linewidth}
\includegraphics[width=1\linewidth]{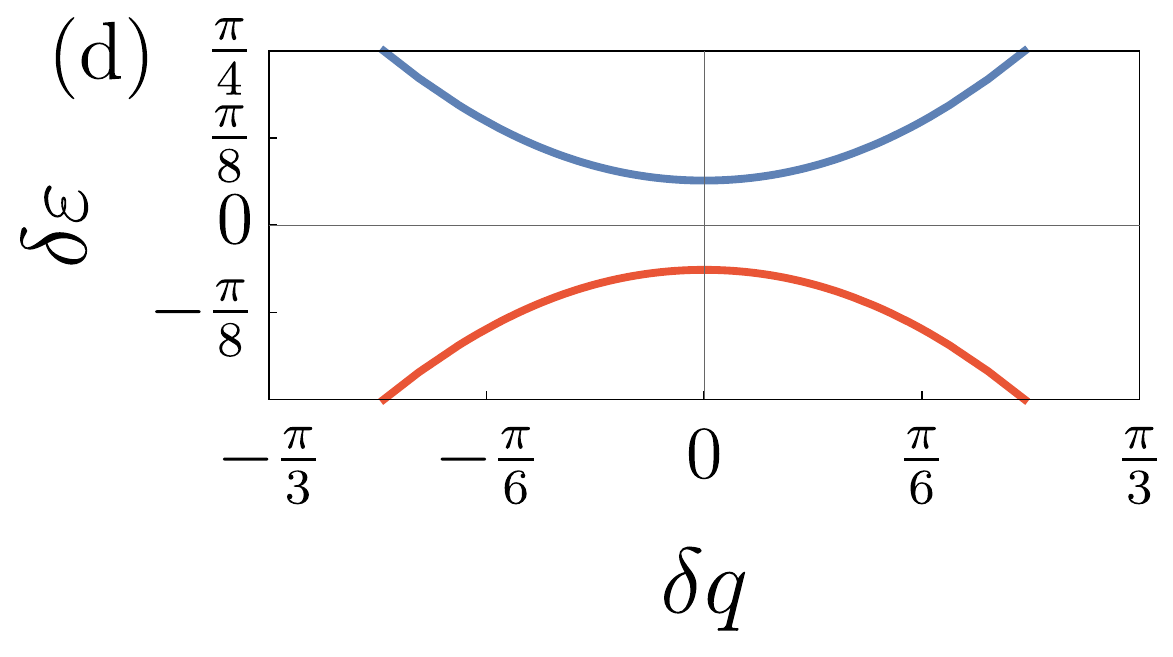} \vfill
\includegraphics[width=1\linewidth]{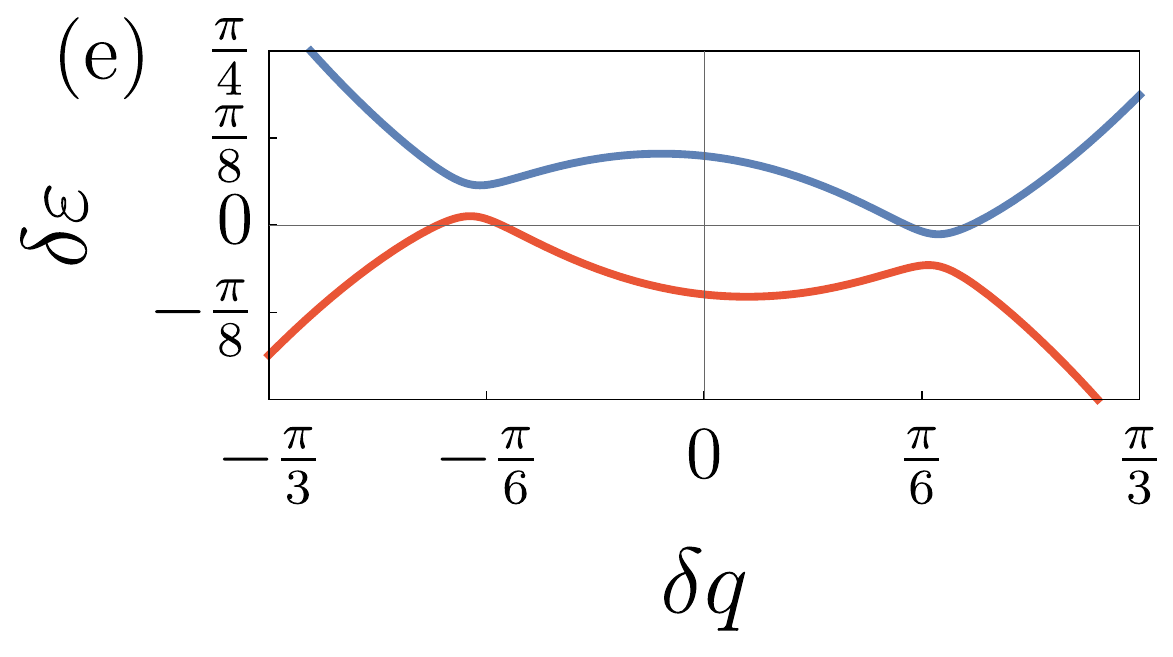} 
\end{minipage} 
\caption{\label{DDP-fig}
 (a) Two close massive  Dirac cones  for each band $\phi=0.2, $ $\beta=0.1,$  and $v=3$ [spectrum is found from Eq.~\eqref{DDP-general}].  Spectrum in vicinity of the convergence point [see Eq.~\eqref{DDPdg}]:  (b) symmetric double DP for negative $\delta \phi$  ($\delta \phi=-0.05,$)   $ \beta=0.05$, and $\delta v =0$; (c) exact convergence of two DPs at $\delta \phi=0,$ $ \beta=0.05,$ and $ \delta v =0$; (d)  disappearance of pair of DPs and increase of gap at positive $\delta \phi$ ($\delta \phi=0.03$), $\beta=0.05$ and $\delta v=0$; (e) asymmetric spectrum for negative $\delta \phi$ and non-zero $\delta v $:     $\delta \phi =-0.06, $ $\beta=0.05,$ and  $\delta v=-0.2.$ 
 Curves in all panels are plotted for     $\alpha=\pi/4.$
 }
\end{figure*}

As  seen from equations presented  in Sec.~\ref{sec:Dirac-points},   different  
massive Dirac cones become closer when flux deviates from integer and half-integer values by a value $\delta \phi$.  For certain values of flux these points  converge forming a double DP  and then disappear with further increase  of   $\delta \phi.$   At the convergence point two branches of the spectrum touch each other, i.e. they coincide and have the same derivative (see Fig.~\ref{DDP-fig}). 

 For simplicity, we limit ourselves below with the  case $\beta \ll 1.$   For $\beta \equiv 0,$ the convergence points appear for four values of the flux, $\phi_j$ ($j=1,2,3,4$), given by:
\be
\begin{aligned}
&\phi_1=1-\phi_2= \frac{ \arcsin (\sin (v/2) \sin\alpha)}{\pi},
\\
&\phi_3=1-\phi_4 =\frac{1}{2}- \frac{ \arcsin (\cos(v/2) \sin\alpha)}{\pi}.
\end{aligned}
\label{phi-alpha}
\ee
Close to these convergence points and for $\beta \ll 1$ the spectrum is approximated as
\be
\delta \varepsilon = D \delta q \pm 
\sqrt{4 \beta^2 \sin^2\alpha +(2 \pi \delta \phi+ E \delta q^2 )^2},
\label{DDP-general}
\ee
where $|\varepsilon - D \delta q| \sim |\delta \phi| \sim \delta q^2 \sim \beta \ll 1. $
Here,    $\delta \phi= \phi -\phi_j,~ \delta q =q- q_j,~\delta \varepsilon =\varepsilon -\varepsilon_j,$ where  $\phi_j$ are given by 
Eq.~\eqref{phi-alpha}.   The positions of convergence points in $(q,\varepsilon)$ plane and formulas for coefficients $D$ and $E$ are presented in  the Appendix~\ref{DDP}. The discussion of some limiting cases can be also found there. The spectrum in vicinity of the convergence points, formation of double DP and disappearance of pairs of DP    are illustrated in Fig.~\ref{DDP-fig}. 

\subsection{Band engineering}
\begin{figure*}
\includegraphics[width=0.8\linewidth]{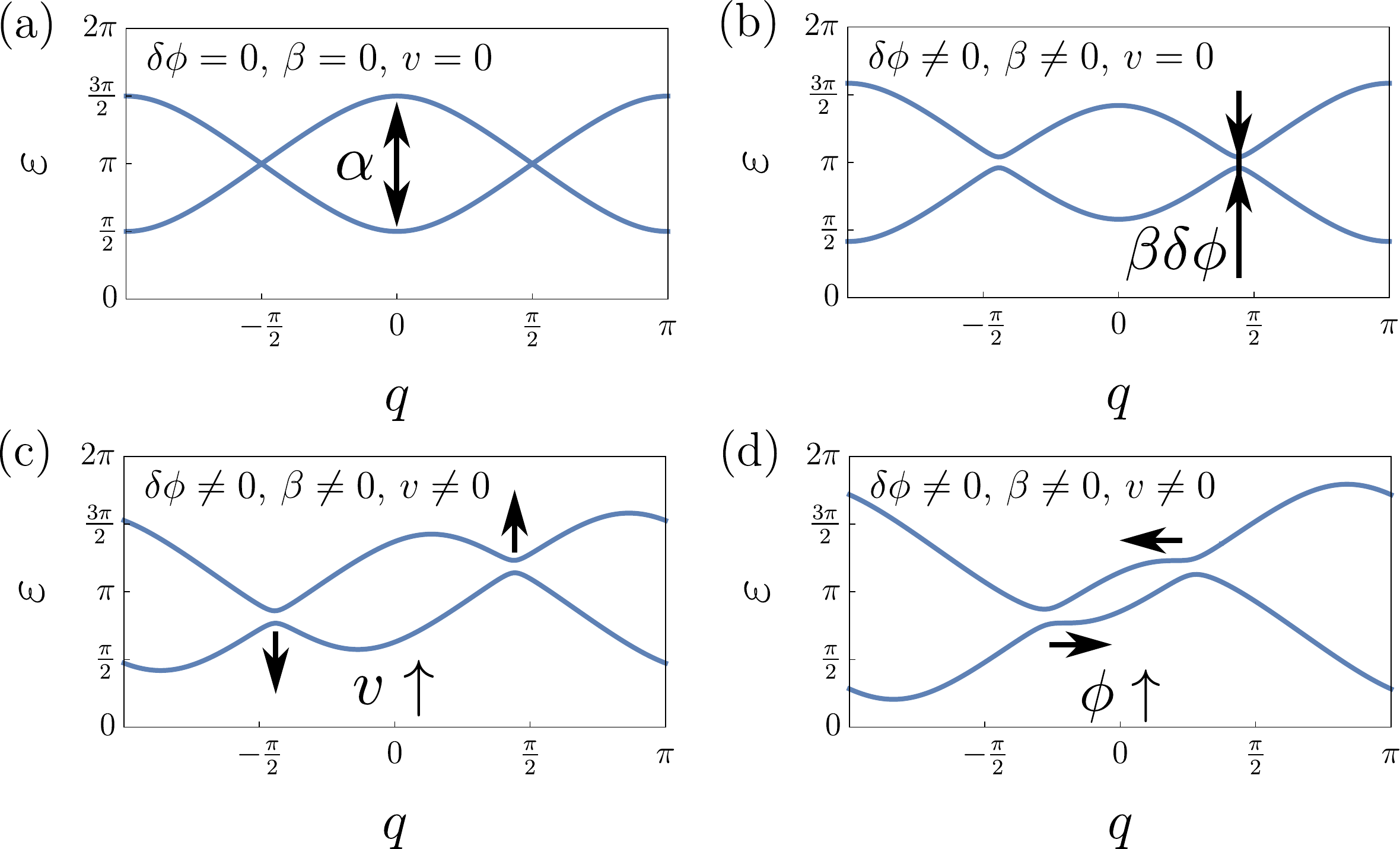}
\caption{\label{fig:sketch}
Schematical picture of the   lowest band  of the crystal  for different values  of parameters: (a)  Two Dirac cones for      $\alpha=\pi/4$,  $\delta\phi\equiv\phi-1/2=0$, $\beta=0$, $v=0$;
(b)  formation of the gap for small deviation of the flux, $\delta \phi$, from half-integer value:   $\delta \phi=0.05$, $\beta=0.6$, $v=0$;
(c) shifting of energy of the DPs in opposite direction for $v\neq 0$: $\delta \phi=0.05$, $\beta=0.6$, $v=\pi/3$;
(d) convergence of  the DPs with increasing of $\delta \phi$: $\delta \phi=0.15$, $\beta=0.2$, $v=\pi/3$. 
}
\end{figure*}

In this section we briefly summarize the discussion of the previous sections in order to describe different ways to control the electron spectrum.  In Fig. ~\ref{fig:sketch} we illustrate the evolution of the bands with a variation of different parameters. First of all, we notice that the  widths of all bands are proportional to $\alpha$  and there is a pair of two symmetric DPs in each band  for special values of the flux  (as illustrated  in Fig. ~\ref{fig:sketch}a for $\phi=1/2.$). The energy positions of the DPs are the same provided $v=0.$ Small deviation of $\phi$ from the value $1/2$ opens small gaps in both Dirac cones (see Fig. ~\ref{fig:sketch}b).  
 The width of appearing gap, i.e.  the mass of the Dirac fermions, is proportional  to $|\delta \phi|$ and tends to zero when $\beta \to 0.$ 
 Changing parameter  $v$ (by using asymmetric gate configuration  shown in Fig.~\ref{fig:nring}a) the energies of the DPs move  in  opposite directions  (see Fig. ~\ref{fig:sketch}c). With increasing $\delta \phi$ two DPs in a pair converge (see Fig. ~\ref{fig:sketch}d). Hence one can control spectrum  using magnetic field and gate electrodes of different configurations.    

\section{
Topologically protected localized states}
\subsection{Effective Hamiltonian near the Dirac point}

Perhaps the easiest way to obtain the effective Hamiltonian near the DPs is to ``take square'' of the dispersion relation. Namely, instead of~\eqref{eq:dispeq} we consider the transfer matrix to the next-to-nearest neighbor, which results in the equation 
\aleq{
\det [\hat 1 e^{i2q} -\hat T^2(\varepsilon)]=0 \, . 
}
To be specific, let us consider  $\phi\simeq 1/2,$  when positions of the  DPs read $q_{\rm DP} = \pm \pi/2 ,$ so that $\hat 1 e^{2iq_{\rm DP}} =-\hat 1.$  One can check that  
 $\hat T^2=-\hat 1$ 
as well for $q=q_{\rm DP}$ (although $T \neq \hat 1 e^{iq_{\rm DP}} $).  This property simplify 
expansion  near the  DPs: $q=\pm \pi /2+ \delta q$, $\varepsilon  =\pi  +\delta \varepsilon$, $\phi= 1/2 + \delta \phi$,  
 $v=\delta v$ ($\delta \phi \ll 1, \delta q \ll 1, ~ \delta v \ll 1$).  We get   
\aleq{
\det \left[ 2i\delta q  + \delta \varepsilon
 \frac{\partial \hat T^2}{\partial \varepsilon}\Big|_{\text{DP}} 
+ \delta v  \frac{\partial \hat T^2}{\partial v}\Big|_{\text{DP}}
+\delta \phi  \frac{\partial \hat T^2}{\partial \phi}\Big|_{\text{DP}} \right]=0 \,.
\label{T-squared}
}
The Hamiltonian with the same dispersion relation can then be found in the form
\aleq{
\mathcal{H}^{(1)}_\text{eff}=-R^{-1} \left.\left( \frac{\partial \hat T^2}{\partial \varepsilon}\right)\right|_{\text{DP}} ^{-1} \left[ 2i\delta q + \left.\frac{\partial \hat T^2}{\partial v}\right|_{\text{DP}}\delta v \right. \\
\left. +\left.\frac{\partial \hat T^2}{\partial \phi}\right|_{\text{DP}}\delta \phi  \right] R \,.
}
Here, $R$ is chosen so that $\mathcal{H}^{(1)}_\text{eff}$ is  hermitian. Taking  \aleq{
R & =
\begin{pmatrix} 
 i \tau ^{\alpha } & -i \tau ^{\alpha}\tau^\beta & i & -i \tau ^{\beta } \\
 \tau ^{\beta } & 1 & \tau ^{\alpha}\tau^\beta & \tau ^{\alpha } \\
 \tau ^{\alpha}\tau^\beta & \tau ^{\alpha } & \tau ^{\beta } & 1 \\
 i & -i \tau ^{\beta } & i \tau ^{\alpha } & -i \tau ^{\alpha }\tau^\beta 
\end{pmatrix} \,,\\ 
\tau^\alpha & = \tan(\tfrac{\pi}{4}+\tfrac{\alpha}{2}) \,, \quad 
\tau^\beta   = \tan{\tfrac{\beta}{2}} \,,\\
}
 after simple rearrangement we obtain the effective Hamiltonian with the  block-diagonal structure
\aleq{
\mathcal{H}^{(1)}_\text{eff} &=
\begin{pmatrix}
\mathcal{H}_1& 0\\
0& \mathcal{H}_2\\
\end{pmatrix} \,, \\
\mathcal{H}_1 & =
\delta v \cos \beta \sin\alpha \sigma_0 +2 \pi \delta \phi \cos \alpha \sin \beta \sigma_1
\\   & 
+ (2 \pi \delta \phi \cos \beta + 2 \delta q \sin \alpha) \sigma_3  \, , \\ 
\mathcal{H}_2 & =
-\delta v \cos \beta \sin\alpha \sigma_0 +2 \pi \delta \phi \cos \alpha \sin \beta \sigma_1
\\ & 
+ (2 \pi \delta \phi \cos \beta  - 2 \delta q \sin \alpha) \sigma_3  \, . 
\label{Heff1}
}
Here, $\sigma_{1,3}$ are the Pauli matrix and $\sigma_0$ is the identity matrix.

Below we will also need corrections to the effective Hamiltonian \eqref{Heff1}.
In order to analyze these corrections, we expand the matrix  $\hat T^2$ in Eq.~\eqref{T-squared}  up to the next order with respect to  small  parameters  $\delta \phi,
~\delta v, ~\delta q, ~\delta \epsilon .$  The calculations are straightforward but cumbersome and will be presented in detail elsewhere~\cite{niyazov-to-be-published}.      
There  are two types of corrections: (i) corrections to  $\bar{ \mathcal{ H}}_1$ and $\bar{ \mathcal{ H}}_2,$ which do not essentially modify the massive Dirac cone spectrum
and can be omitted for our purposes  and  (ii) the block-off-diagonal  corrections which are given by 
       \begin{equation}
    \mathcal{H}^{(2)}_\text{eff} =
\begin{pmatrix}
0& \hat V  \\
\hat V ^\dagger& 0\\
\end{pmatrix} \,.\\
\end{equation}
With the precision necessary for Sec.\ \ref{sec:TunQub} below one finds
\aleq{
\hat V &= 
 i (\bar q^2 -  \pi^2 \delta\phi^2 \sin^2 \beta) \sin 2 \alpha\,\hat \sigma_3 
 \\
 &+  \pi^2 \delta \phi^2 \sin \alpha \sin 2 \beta \,\hat \sigma_1,
}
where $\hat \sigma_i$  ($i=0,1,2,3$) are the Pauli matrices and $\bar q$ is given by Eq.~\eqref{barq}.

\subsection{Topologically protected localized states}
The block $\mathcal{H}_2$  of the Hamiltonian can be obtained from  $\mathcal{H}_1$ by  changing  $\delta q \to -\delta q$ and $\delta v \to -\delta v$ (or $\alpha \to -\alpha$).
The term with $\delta v$ may be  absorbed into an energy shift in each of two blocks:
\aleq{
  E_1&= - E_2= -W \,, \\
W&= \delta v \cos \beta \sin \alpha \,.
\label{W}
}
 So we may  discuss the block $\mathcal{H}_1 $ only.  
 The small wave vector shift, $\delta q \ll 1 $, can be represented  as a electron's motion along the rings' array,  $\delta q =-i d/dx,$  where $x$ is the coordinate  along the array measured  in  units of $L.$     
 Making rescaling of this dimensionless coordinate:  $ {2} \sin \alpha \bar x = x ,$ we find   
  ${2} \sin \alpha \delta q 
  =
  -i   \tfrac{\partial}{\partial \bar x}$.    Finally, $\bar{\mathcal{H}}_1 $ becomes a Hamiltonian of massive Dirac particle: 
\aleq{
\bar{\mathcal{H}}_1=
\begin{pmatrix}
\bar \phi - i \frac{\partial}{\partial {\bar x}} & h \\
h & -\bar \phi + i \frac{\partial}{\partial {\bar x}}\\ 
\end{pmatrix} \,,
}
where $\bar \phi = 2\pi \delta \phi \cos \beta$ and $h= 2\pi \delta \phi \sin \beta \cos \alpha$. The dispersion of such Hamiltonian for a plane wave, $e^{i\bar q {\bar x}}$ is given by  $E^2=(\bar q+\bar \phi)^2+ h^2,$ where  \be \bar q= 2  \sin \alpha ~\delta q 
=-i\frac{d}{d \bar x} \,.
\label{barq}
\ee

Let us now discuss localized states near the defects. We assume that there are two half-infinite regions of the array with different value of the flux. In the first region, at $x<0$, the flux is smaller than $1/2$, $\phi=1/2-\delta \phi$, whereas in the second one, at $x>0$,  we have $\phi=1/2+\delta \phi$. 
The problem is fully analogous to the one discussed in Ref.~\cite{Volkov1985}. Then at the interface between two regions there appear  two (because Hamiltonian has two blocks)  degenerate localized states of the Volkov-Pankratov type:
\aleq{
\psi^{\text{loc}}_1= 
\begin{pmatrix}
i \\
1 \\
0\\
0\\
\end{pmatrix}
e^{-(|h| + i|\bar \phi |)|{\bar x}|}, \,  \\
\psi^{\text{loc}}_2= 
\begin{pmatrix}
0\\
0 \\
1\\
i\\
\end{pmatrix}
e^{-(|h| -i|\bar \phi |)|{\bar x}|} \, 
\label{localized}
}
 Note that we removed the prefactor $(-1)^{n}=\exp( \pm i \pi x )$ from both functions, which corresponds to the shift  $\delta q= q  \pm \pi/2$ mentioned above.
We see that the localization lengths of these localized states are equal. For $\delta v=0,$ the energy corresponding to these degenerate states lies exactly in the middle of the Dirac gap just as in the problem discussed in    Ref.~\cite{Volkov1985}.

The localized topologically  protected states obtained above form a two level system, or qubit.  The energy levels of this system  
depend on the parameter $\delta v$ which controls up-down asymmetry of the system.  
Changing $\delta v$ by gate electrodes one can manipulate  energy levels $E_1 $ and $E_2 $ which cross each other,  for  $\delta v=0,$ see Eq.\eqref{W}. At this point, i.e.\ for the  setup with up-down symmetry, the levels are doubly degenerate within the lowest approximation with respect to $\delta \phi$, $\delta q.$

\subsection{Tunable topologically protected qubit} 
\label{sec:TunQub}
\begin{figure}
 \includegraphics[width=0.7\columnwidth]{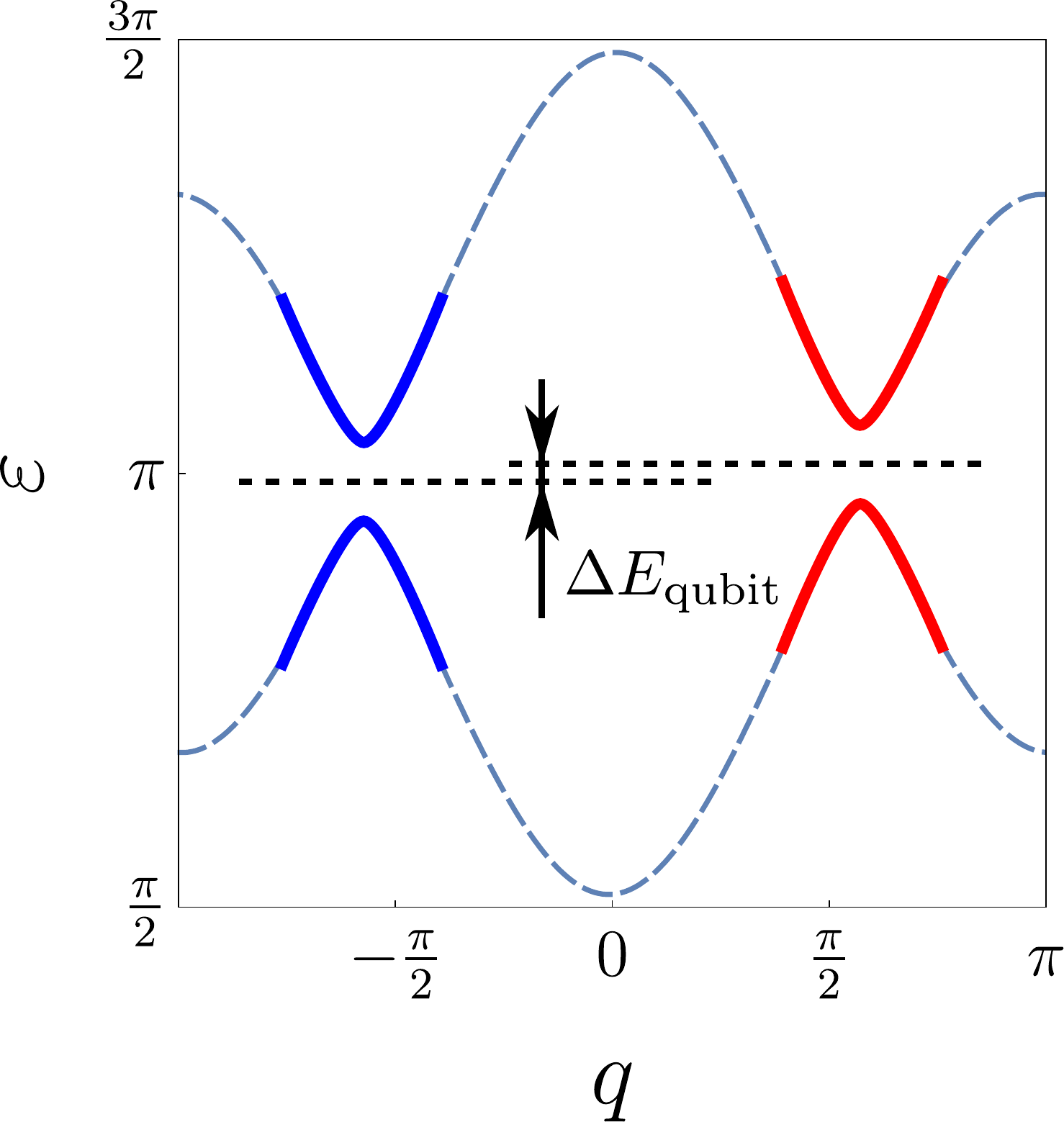}
\caption{\label{fig:dispersion}
Dispersion of the lowest band of the  helical crystal for $(\alpha=\pi/5,\beta=0.6,v=0.07, \phi=0.45)$. The vertical distance between two DPs is  controlled by $v.$   Dashed lines shows positions of localized levels arising when the crystal has a defect. $\Delta E_{\rm qubit}=E^+_{\rm qubit} -E^-_{\rm qubit}$, see Eq.~\eqref{Epm-qubit}.   }
\end{figure}

Projecting $H_{\rm eff}^{(1)}+H_{\rm eff}^{(2)}$ onto two localized states, Eq.~\eqref{localized}, we get the Hamiltonian  of the qubit formed by $\psi_1^{\rm loc}$ and $\psi_2^{\rm loc}$:
\begin{equation}
 \mathcal{H}_{\rm qubit} =
\begin{pmatrix} -W & \Delta_{\rm qubit}  \\
\Delta_{\rm qubit}^*  &  W
\end{pmatrix} \,,
\label{Hq}
\end{equation}
Here, $W$ depends linearly on $\delta v$ according to  Eq.~\eqref{W}, 
$\Delta_{\rm qubit} = \Delta_x+ i \Delta_y= 2\pi \delta \phi^2 \cot \alpha \sin\beta (\cos \alpha \cos \beta +  i \sin\beta).$ 
Corresponding energy levels are given by 
\be
 E^{\pm}_{\rm qubit}= \pm \sqrt{W^2 +  |\Delta_{\rm  qubit}|^2  },
\label{Epm-qubit}
\ee
where 
  \be 
|\Delta_{\rm qubit}|= 2\pi^2 \delta \phi^2 
\sin\beta \cos\alpha \sqrt{\cot^2\alpha + \sin^2 \beta}, 
\label{D-qubit}
\ee  
see Fig.~\ref{fig:dispersion}.

The qubit is equivalent to a spin-1/2 in the  ``magnetic field'' 
\be
\mathbf B_{\rm qubit} = (\Delta_x,\Delta_y, W).
\label{B-qubit}
\ee
Most importantly, the ``magnetic field'' $\mathbf B_{\rm qubit}$   actually depends on the gate voltages and can be controlled in a pure electrical way. We notice also that such qubits exist in all bands, so that for high temperature there is an  ensemble  of  $T L/v_{\rm F}$ qubits in the temperature window. These qubits can be coherently manipulated by gates and magnetic field.

\section{Effect of  the electron-electron interaction}
\label{interaction}

Next, we discuss effects of the electron-electron interaction  on the properties of the crystal.  It is known that the  interaction strongly renormalizes the properties of the contact between two helical states  \cite{Teo2009,Aristov2016}, including spin asymmetry of the tunneling \cite{Aristov2017}.
Here, we use renormalization group (RG) equations derived in  Ref.~\cite{Aristov2016}  to find a general phase diagram of the helical crystal. 

The problem of renormalization of the helical crystal is two-fold. One aspect is the renormalization of the individual contact, which was discussed in \cite{Aristov2016}. Let us discuss now the second aspect which is the peculiarity of the crystal geometry that restricts the renormalization by certain scale. 

There are several   relevant  length scales in the problem: the Fermi wavelength   $\lambda_{F} $ (ultraviolet cut off), $ l_{T}=v_{\rm F}/T $ (thermal length),  and    ring size  $L.$ 
Corresponding energy scales are given by
$E_{\rm F},$ $T,$ and   $v_F/L$ (level spacing in a single ring). 
We assume that   
$E_{\rm F} \gg v_{\rm F}/L$  
  and $E_{\rm F} \gg T.$   
The relation between $T$  and  $v_{\rm F}/L$  can be arbitrary.

As discussed to some detail in Appendix \ref{sec:AppRen}, 
we propose the following scenario. 
 The renormalization of separate junctions happens at relatively high temperatures when the running energy scale is    
\be E_{\rm F}> E >\max [T,v_{\rm F}/L] ,
\label{interval}
\ee (or, equivalently, the spatial interval lies between $\lambda_{\rm F}$ and   $\min [L, l_T]   $). When the temperature becomes low, $T<v_{\rm F}/L,$ the junctions cannot be regarded as separate from each other and form an effective media. In other words, RG flow for the junctions' renormalization stops when  $T$ decreases below $v_{\rm F}/L.$    

Hence, interval ~\eqref{interval}  corresponds to independent renormalization of  separate junctions, {which} is the case discussed in Ref.~\cite{Aristov2016}, see also \footnote{The renormalization was discussed  in \cite{Aristov2016}
 in terms  of dimensionless reduced conductances through a single junction,
  $G_{\rm R}=(1-a)/2$ and $G_{\rm D}=(1-b)/2,$  which are connected  to the angles $\alpha,~\beta$ ($\beta$ and $\gamma$ in terms of~\cite{Aristov2016}, correspondingly) as follows:
$ a=2 \cos^2 \beta \sin^2 \alpha-1, \quad b=\cos 2 \alpha$   
}.
Using Eq.~(18) of Ref.~\cite{Aristov2016} and assuming that the dimensionless interaction strength, $g$, in the helical state is small (so that the Luttinger parameter $K=[(1-g)/(1+g]^{1/2}$ is close to unity, $K \simeq 1-g$) we find RG flow in terms  of  $\alpha,~\beta$: 
\aleq{
\frac{d \alpha}{d \Lambda}&= -\frac{1}{8} g^2 \sin 2 \alpha  \left[2-\sin ^2\alpha  \left(4-\sin ^2 2 \beta \right)\right] \,, \\
\frac{d \beta}{d \Lambda}&= -\frac{1}{8} g^2 \sin ^2\alpha \sin 4 \beta \,
. \label{RG}} 
Here $\Lambda = \log E_{\rm F}/E.$ We note that parameter $v$ does not flow.   

\begin{figure}
\includegraphics[width=0.95\linewidth]{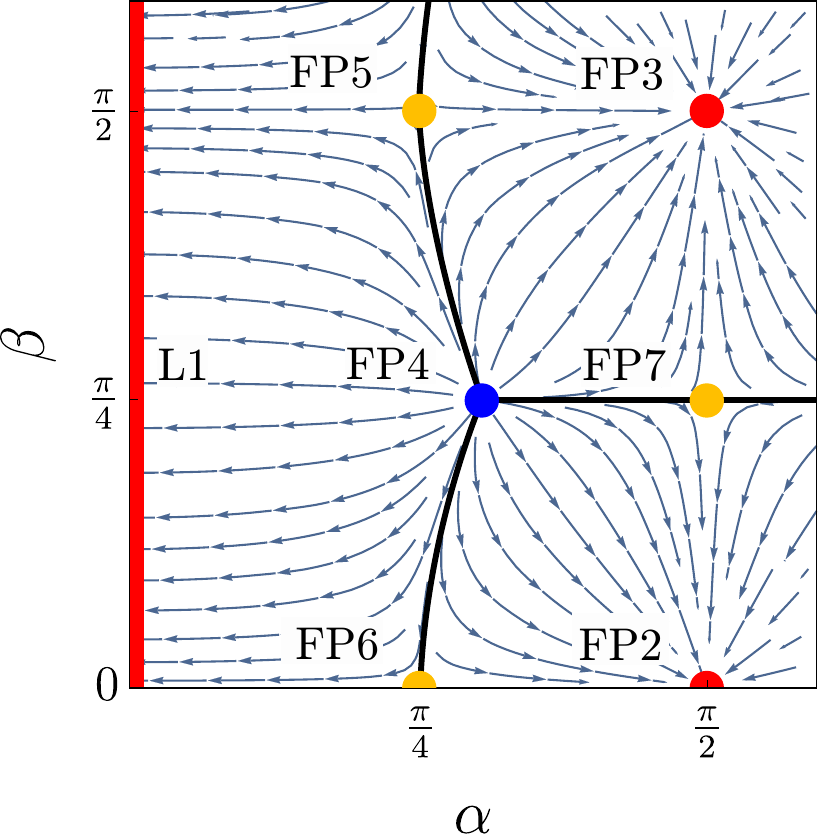}
\caption{\label{fig:RGflows}
Phase diagram and RG flows at the plane $(\alpha,\beta)$ [equivalent to  Fig. 2a of Ref.~\cite{Aristov2016} with corresponding change of parameters]  
Red color ---  the line of  stable fixed points (L1) and two stable fixed points FP2 and FP3;  blue color---fully unstable {\it multicritical}  fixed point, FP4,  separating three stable phases, corresponding  to L1 (independent rings),   FP2 (independent  shoulders), and FP3  (spin-flip chiral channels);    
Yellow color --- three unstable saddle fixed point, FP5, FP6, and FP7.  
All fixed points are listed in the Table \ref{tab:fp}.  }
\end{figure}

This set of equations defines the phase diagram  of the RG flow in the plane $(\alpha,\beta)$ as shown in Fig.~\ref{fig:RGflows}(this diagram is equivalent to  Fig. 2(a)  of Ref.~\cite{Aristov2016} with corresponding change of variables). 
The diagram contains
a line of fixed points L1 
and several fixed points FP2-FP7   
shown by different colors: Red color ---  the line of  stable fixed points L1, corresponding to $ \alpha=0,$ and two stable fixed points  FP2=$(\pi/2,0)$ and FP3$=(\pi/2,\pi/2)$;   blue color---fully unstable fixed point   
FP4$=(\arctan\sqrt 2,\pi/4);$ 
 yellow color --- three unstable saddle fixed points, FP5$=(\pi/4,\pi/2)$,  FP6$=(\pi/4,0)$,    
  FP7$=(\pi/2,\pi/4).$ All FPs are listed in the Table ~\ref{tab:fp}. 

We see that there are three stable phases in the crystal described by  stable line L1,  and  stable points FP2 and FP3. Physics behind  these phases is as follows.   
 Stable line  L1 corresponds to {\it independent rings}, when tunneling coupling between rings is zero and bands of the crystal shrink  into discrete levels, Eq.~\eqref{levels}.   Stable point FP2 corresponds to {\it independent  shoulders} of the rings array. Indeed, at this point we have $t=f=0,~r=1.$  As seen from Fig.~\ref{rtf}, in this case upper and lower shoulders of the array are disconnected and electron moves without changing direction and chirality while passing contacts.    Finally,  FP3  corresponds to {\it spin-flip chiral channels}.  At this point, $t=r=0,~f=1.$ This means (see Fig.~\ref{rtf}) that electron makes spin-flip  after each tunneling jump but does not change direction of the propagation.       

Very interesting feature of the  RG phase diagram is {\it multicritical } behaviour. The  fully unstable point FP4 separates  three stable phases described above.

One can show that this phase diagram remains  unchanged when taking into account  higher order contributions in $g$ \cite{Aristov2016}.

\begin{table}
\centering
\begin{tabular}{|c| c | c |c| c| c| c| c|} 
\hline
\textnumero&L1&FP2&FP3&FP4&FP5&FP6&FP7\\
 \hline
 $\alpha$ & 0 & $\pi/2$ & $\pi/2$ & $\arctan \sqrt{2}$ & $\pi/4 $& $\pi/4 $& $\pi/2$ \\ 
 $\beta$ & arbitrary & 0 & $\pi/2$ & $\pi/4 $& $\pi/2 $& 0 & $\pi/4$  \\
 \hline
 $t$ & 1 & 0 & 0 & $1/\sqrt{3}$& $1/\sqrt{2}$ & $1/\sqrt{2}$ &  0 \\
 $r$ & 0 & 1 & 0 & $1/\sqrt{3}$& 0 & $1/\sqrt{2}$ &  $1/\sqrt{2}$ \\
 $f$ & 0 & 0 & 1 & $1/\sqrt{3}$& $1/\sqrt{2}$ & 0&  $1/\sqrt{2}$ \\ 
 \hline
 Stability & s & s & s & u & u & u & u \\ 
  \hline
\end{tabular}
\caption{ Line  of stable fixed points  (L1) and  fixed points (FP2-FP7)  of interaction-induced  RG flow in the plane $(\alpha,\beta).$ Stability is denoted by $s$ (stable) or $u$ (unstable).
}
\label{tab:fp}
\end{table}

\section{Outlook and conclusion}
Before concluding  the paper let us outlook  interesting problems for future research.   
As we demonstrated, two localized states, which appear on the boundary between regions  having  different   radii  of rings  form a qubit placed in the effective ``magnetic field''  $B_{\rm qubit}$ [see Eq.~\eqref{B-qubit}].    Actually, this field depends  on gate voltages, so that the state of qubit can be manipulated purely electrically. Fabricating crystal with several boundaries between  different regions allows one to create  spatially separated qubits.    By changing  $B_{\rm qubit}$ (adiabatically or abruptly)  one can study  single qubit evolution  and creation of  entangled states involving      qubits placed at different boundaries.  Also,  even at one boundary there can be  many qubits for high temperatures, when temperature window covers many bands of the crystal and, consequently, involves  many  massive Dirac cones.      Related  tasks are electron-phonon dephasing and optical excitation  of the ensemble of these  spatially and/or energy separated qubits (analysis  of quantum computing by an ensemble of energy separated qubits in a AB interferometer was presented in Ref.~\cite{Niyazov2020}).    
    
    There are also several  directions of improvement  of the  developed model.    In the above calculations of  energies $E^\pm_{\rm qubit}$    we assumed that both states forming qubit are inside the  Dirac gap:   $E^+_{\rm qubit} - E^-_{\rm qubit} < \Delta.$ This inequality  fails with increasing $v.$ Study of the corresponding case  of  sufficiently large $v$ implies  analysis of the decay of qubit states into continuous spectrum. Another problem is related to the interaction-induced  renormalization of the qubit parameters. Indeed, above we argued that RG flow stops with decreasing the temperature below level spacing in a single ring. This  conclusion was based on  the homogeneity of the interacting  crystal. It fails  in the presence of a defect. Hence,  qubit parameters can flow  even  at very low temperature. Note finally, that additional physics might appear for special models of edge confining potential, where edge reconstruction is possible~\cite{Wang2017a}.     
    
    Analysis of all these interesting problems  is out of scope of the current paper and will be presented elsewhere.                                  
To conclude,  we developed a theory of tunable helical crystal 
 formed by  periodic array of    identical holes in a  2D topological insulator.   We demonstrated that the tunneling transport through such array can be  controlled  by  magnetic field and also by  purely electrical way using various configuration of gate electrodes. For integer and half-integer values of the  dimensionless flux through the holes, pairs of the  { Dirac}  points appear in each band of the spectrum. The positions of these points and  
 the    velocities at these points  can be tuned by the gate voltages.  Deviation of the flux from the special values leads to appearance of gap, i.e. corresponding Dirac cones become massive, and to convergence of the different Dirac points.        
 The  electron-electron interaction renormalizes properties of the crystal. We find the phase diagram and RG flows of the system  in the plane of parameters $(\alpha,\beta)$ characterizing total and spin-flip tunneling  couplings, respectively. 
 We find that there are  three different phases: independent rings, independent shoulders,  and perfect spin-flip channels.  There exist a fully unstable multicritical fixed point separating these phases. 
 
   We also derive an effective Hamiltonian of a    defect  in the crystal and show that topologically protected    qubits can appear in the crystal within the gaps of the massive   Dirac cones. These qubits  can be controlled by purely electrical way.  At relatively large temperatures,  $T\gg v_{\rm F}/L,$ an ensemble of  $N \sim T L/v_{\rm F} $ topologically protected qubits appears.   
   The possibility of purely  electrical high-temperature control of these qubits opens a wide avenue for applications in the area of quantum computing.

\section*{Acknowledgements}

The work was funded by 
the Russian Science Foundation, Grant No. 20-12-00147-$\Pi$.  The work of R.A.N. was partially supported by the Theoretical Physics and Mathematics Advancement Foundation ``BASIS''. Investigation  of double Dirac points    (Sec.~\ref{main-DDP}) was partially supported by a grant of the President of the Russian Federation for state support of young Russian scientists - Candidates of Science (project No. MK-2918.2022.1.2,   support of R.A.N.).

\appendix
\section{Transfer matrix } \label{app:transfer}

The transfer matrix $\hat T_0,$ which connects  amplitudes  on the left and right sides of a contact is defined as follows        \aleq{
&\begin{pmatrix}
C_3^\prime \\
C_3\\
C_4^\prime\\
C_4 \\
\end{pmatrix}_{\!\!\!\!(n+1)}
=
\hat T_0
\begin{pmatrix}
C_2 \\
C_2^\prime \\
C_1\\
C_1^\prime \\
\end{pmatrix}_{\!\!\!\!(n)} \, .
}
This definition allows one to find $\hat T_0$ by  using   matrix $\hat S_0$ [see Eq.~\eqref{eq:smat}]. Direct calculation shows that  $\hat T_0$ is given by  Eq.~\eqref{T0}.

While propagating inside a  certain ring, the electron waves acquire phases that depend  on electron energy as described by matrix  Eq.~\eqref{def:Pmat} which connects 
amplitudes inside the ring as follows
\be
\begin{pmatrix}
C_2 \\
C_2^\prime \\
C_1\\
C_1^\prime 
\end{pmatrix}_{\!\!\!\!(n)}=\hat P_0 
\begin{pmatrix}
C_3^\prime \\
C_3 \\
C_4^\prime\\
C_4 
\end{pmatrix}_{\!\!\!\!(n)}.
\ee

By using gate electrode shown in Fig.~\ref{fig:nring} one can make  different the energies for up and down part of the ring.      Then matrix $\hat P_0$ becomes 
\aleq{
\hat P_0=\text{diag} [e^{i \varphi_b^{\rm up}},e^{-i \varphi_a^{\rm up}},e^{i \varphi_a^{\rm down}},e^{-i \varphi_b^{\rm down}}]\, ,
\label{def:Pmat-general}
}
where
\be
\begin{aligned}
  &\varphi_a^{\rm up}= \varepsilon^{\rm up}/2  -\pi \phi,  \\
  &\varphi_b^{\rm up}= \varepsilon^{\rm up}/2  +\pi \phi,
  \\
  & \varphi_a^{\rm down}= \varepsilon^{\rm down}/2  -\pi \phi,
  \\
  &\varphi_b^{\rm down}= \varepsilon^{\rm down}/2  +\pi \phi,
\end{aligned}
\label{phases-general}
\ee
and $\varepsilon^{\rm up,down}= k^{\rm up,down} L.$
The meaning   of these phases is the following.   ``Up'' and ``down'' stand for the  upper and lower  shoulders of the ring. We assume that   the upper  and lower  energies can be different in general by application of the different gate voltages to upper and lower shoulders of the ring,  so that the kinematic phase $\varepsilon^{\rm up, down }  = k^{\rm up, down} L$ can  be different, too.
Equations \eqref{phases} are reproduced from Eq.~\eqref{phases-general} when
$\varepsilon^{\rm up}=\varepsilon^{\rm down}=\varepsilon.$

\section{Phases in S-matrix  and transfer matrix
}

Observing symmetry with respect to time  inversion, we should impose the following condition $S^T= - E S E$ with $E=\mbox{diag}[1,-1,1,-1]$. 
This condition allows one to write the $S$ matrix in the most general form 
given by Eq.~\eqref{eq:smat}. 
Corresponding  transfer matrix is given by
\aleq{
\hat T_0 =& \frac{e^{-i(\gamma_1+\gamma_2)}}{1-t^2}\\
\times& \begin{pmatrix}
r_2 e^{i(\gamma_1+\gamma_2)}& f_1 t_2 & -f_1 e^{i(\gamma_1+\gamma_2)} & -r_2 t_2\\
-t_1 f_2  & -r_1& r_1 t_1& f_2  \\
-f_2 e^{i(\gamma_1+\gamma_2)}&-  r_1 t_2 & r_1 e^{i(\gamma_1+\gamma_2)} & f_2 t_2\\
t_1 r_2 & f_1& -f_1  t_1 & -r_2  \\
\end{pmatrix}
}
Let us now  assume the parametrization 
\[
\begin{aligned}
\tau_1=(\gamma_1+\gamma_2)/2+\delta \tau \,,\\
\tau_2=(\gamma_1+\gamma_2)/2-\delta \tau \,,\\
\varphi_1=(\gamma_1+\gamma_2)/2+\delta \varphi \,,\\
\varphi_2=(\gamma_1+\gamma_2)/2-\delta \varphi\,.\\
\end{aligned}
\]
We notice that all phases in $\hat  T_0$ can be removed by the following operation 
\[
\begin{aligned}
T_0   = & \hat D_1\hat T_{r}   \hat D_2  \,, \\    
 \hat D_1  = &\mbox{diag}[ e^{i \gamma_2/2}, e^{ i (\delta \tau-\delta \varphi-\gamma_2/2)}, 
 e^{i(\gamma_1/2-\delta \varphi)}, e^{ i(\delta\tau -\gamma_1/2)} ] \,,\\
\hat D_2  = &\mbox{diag}[ e^{i\gamma_2/2}, e^{i (-\delta \tau+\delta \varphi-\gamma_2/2)}, 
e^{i(\gamma_1/2+\delta \varphi)}, e^{i(-\delta\tau -\gamma_1/2)} ] \,, \\ 
\end{aligned}
\]
where $\hat T_{r} $ is given by above $\hat  T_0$ with  $\gamma_1, \gamma_2, \delta\tau ,\delta\varphi $ put to zero. 
One can show, that phases
$\delta \tau,\delta \varphi$ do not enter dispersion equation for the spectrum of the crystal. Putting   $\delta \tau=\delta \varphi=0,$ we arrive at Eq.~\eqref{T0} of the main text.

\section{Double   Dirac points  }\label{DDP}

The positions of convergence points $(q_j,\varepsilon_j)$ read
\be
\begin{aligned}
&(q_1,\varepsilon_1) =(q_2,\varepsilon_2)=(\pi/2,~2\pi+ 4\pi n) ~ {\rm and} 
\\
&(q_1,\varepsilon_1)=(q_2,\varepsilon_2)=(-\pi/2,~  4\pi  n);
\\
&(q_3,\varepsilon_3) =(0,-\pi+ 4\pi n) ~ {\rm and}~(\pm \pi,\pi+ 4\pi n); 
\\
&(q_4,\varepsilon_4) =(0,\pi+ 4\pi n) ~ {\rm and}~(\pm \pi,-\pi+ 4\pi n).
\end{aligned}
\ee
Here $n$ is integer. Coefficients $D$ and $E$ also depend on $j,$  $D= D_j,$ $E= E_j,$ where
\be
\begin{aligned}
& D_2=-D_1= \frac{2 \sin \alpha \cos(v/2)}{\sqrt{1- \sin^2(v/2) \sin^2 \alpha}},
\\
&E_1=-E_2=\frac{\cos^2\alpha \sin \alpha \sin (v/2) }{(1- \sin^2(v/2) \sin^2 \alpha)^{3/2}},
\\
& D_4=-D_3= \frac{2 \sin \alpha \sin(v/2)}{\sqrt{1- \cos^2(v/2) \sin^2 \alpha}},
\\
&E_4=-E_3=\frac{\cos^2\alpha \sin \alpha \cos (v/2) }{(1- \cos^2(v/2) \sin^2 \alpha)^{3/2}}.
\end{aligned}
\ee

Above equations simplify for $v$ close to $0$ and $\pi.$

For example, let us assume that   
\be v= \delta v \ll 1 ,   \ee
and consider  point corresponding to  $j =4.$  We get
\be
\begin{aligned}
&\phi_4 \approx \frac{1}{2} +\frac{\alpha}{\pi} - \frac{ \tan\alpha }{8 \pi}~ \delta v^2,
\\
&D_4 \approx -2\delta v \tan \alpha ,~E_4\approx \tan \alpha. 
\end{aligned}
\ee
Expression for  spectrum becomes:
\be
\begin{aligned}
& \delta \varepsilon \approx  \delta q\delta v \tan\alpha  
\\
&
\pm \sqrt{4\beta^2 \sin^2\alpha+ [ \delta q^2\tan\alpha +2\pi \delta\phi]^2}. 
\end{aligned}
\label{DDPdg}
\ee
Here, we assumed that 
\be
\beta \sim \delta \varepsilon \sim \delta \phi \sim \delta v^2 \sim \delta q^2 \ll 1. \label{ineqDDP}  
\ee
The spectrum simplifies even more and becomes symmetric with respect to $\delta q$ for $\delta v =0$:  
\be
\delta \varepsilon = \pm \sqrt{4\beta^2 \sin^2\alpha+ ( \delta q^2\tan\alpha +2\pi \delta\phi)^2} .
\label{DDPdg0}
\ee
As seen from Eq,~\eqref{DDPdg0},  for $\delta \phi<0,$ the spectrum has two close DP separated  by distance $\delta q \propto \sqrt{-\delta \phi}.$ Exactly at $\delta\phi=0,$ the DP converge. In this case, spectrum has  a small gap $\Delta=4 \beta \sin \alpha $  at $\delta q=0,$         increases in a  quatric way   at small $\delta q,$ and becomes parabolic at large $\delta q$ with an effective  ``mass'' depending on tunneling coupling only:
\be
\delta \varepsilon = \pm \sqrt{4\beta^2 \sin^2\alpha+  \delta q^4\tan^2\alpha } .
\label{DDPdg0df0}
\ee
For positive $\delta \phi,$ the gap increases, $$\Delta \to 4
\sqrt{\beta^2 \sin^2\alpha+   \pi^2 \delta\phi^2},$$   
and spectrum, Eq.~\eqref{DDPdg0}, becomes parabolic both  at very  small $\delta q$ and very large  $\delta q,$ but with different effective masses. 
These cases are illustrated in Fig.~\ref{DDP-fig} in panels (a,b,c). We also demonstrate in this figure [see panel (d)]  asymmetric double DP spectrum corresponding to $\delta v \neq 0$ and $\delta \phi<0.$

One more simple  example of double DP is realized for $v $ close to  $\pi$  at the point corresponding to  $j=2.$
Then,  $\phi_2 \approx 1- \alpha/\pi + \delta v^2 ~\tan \alpha/8\pi, $ $D_2= -\delta v \tan \alpha,$ $E_2= -\tan \alpha.$ The spectrum is described by  Eq.~\eqref{DDPdg} with the replacement $\delta v\to -\delta v$ and  $\delta \phi \to -\delta \phi. $ 

\section{Renormalization of helical crystal}
\label{sec:AppRen}

In this Appendix, we justify the choice of the infrared cut-off in Eq.~\eqref{interval}.   Usually, e.g. in a single tunnel junction, renormalization  arises due to virtual transitions involving  energies    $E$ (counted from the Fermi energy) within the  energy interval~\cite{schulz2000fermi,GiamarchiBook} 
\be
 E_{\rm F}> E> T,
\label{RG-int}
\ee
corresponding to  the  spatial scales between $\lambda_{\rm F}$ and $l_T.$  

In a helical crystal  the situation is more subtle and the choice of the infrared cut-off deserves a special discussion. The integration over all possible states within the interval \eqref{RG-int}  can be represented as a summation over  crystal band indices (stemming from single ring levels) and integration over wave-vector $q.$  
Two limits  $v_{\rm F}/L \gg T$ and    $v_{\rm F}/L \ll T$ correspond to different physical situations (for discussion of similar problem of resonance defect formed by the two barriers in the Luttinger liquid separated by distance $L$ see Ref.~\cite{Polyakov2003}).

In a simpler case of relatively high temperatures, $T \gg v_{\rm F}/L,$  the thermal length is shorter than $L.$ 
 Since the correlations between electrons are lost beyond the thermal length, the  different junctions  are renormalized separately,  
 with $l_T$ being the infrared cut off.  In this case,  the temperature is high compared to characteristic  band width, and $\sim v_{\rm F}/L,$  the  specific structure of the bands is irrelevant. 

At lower temperatures, when $l_T \gg L$, the  interval  \eqref{RG-int} can be split  into two regions: $ v_{\rm F}/L < E < E_{\rm F}$  and $ T<E<v_{\rm F}/L.$   In the first region, we probe the scales shorter than $L,$   so that one can consider renormalization of individual junctions.  
By contrast, in the second region,  corresponding spatial scales are larger than $L$  and        RG flow  should be discussed in terms  of the crystal  bands. More specifically, the Fermi level ``picks''   
a certain band number $N$  and a  point in the 
dispersion curve corresponding to a certain $q,$  and renormalization is determined by   energies  lying in this and neighboring bands.      
Individual rings and junctions play a role of building blocks of the effective uniform media. The consideration of  phenomena happening at such  scales cannot be described in terms of separate junctions. Instead, one  starts with a consideration of interacting crystal with a certain dispersion curve, $E_N(q).$   
The next step is to linearize the spectrum near the 
Fermi level $E_{\rm F}$ corresponding to $q_{\rm F},$ which obeys $E_{\rm F}=E_N(q_{\rm F}).$
The linearized spectrum is a starting point for the description in terms of the Luttinger liquid.  The parameters of such \emph{homogeneous} liquid 
depend on the interaction strength in a trivial manner, they are either of marginal scaling dimension 
(in case as Fermi velocity)  or  irrelevant  (dispersion curvature)~\cite{schulz2000fermi,GiamarchiBook}.     The non-trivial renormalization occurs in this lowest energy region only if we have a defect or distinct junction in this liquid. Particularly it implies that the properties of the above topologically protected qubit can be renormalized, which is an interesting problem on its own and should be discussed separately.       
Meanwhile we notice that the parameters of the  energy band $E_N(q)$ are determined by those already renormalized parameters  $\alpha, \beta$, which are obtained by integrating out  the running energies within the first region, $E> v_{\rm F}/L$.

\end{document}